\documentclass[12pt]{article}
\pdfoutput=1

\def\be{\begin{equation}}
\def\ee{\end{equation}}

\usepackage{graphicx}
\usepackage{amsmath,amssymb}
\usepackage{multirow}
\usepackage{subfigure}

\usepackage{cite}

\begin{document}
\titlepage
\begin{flushright}
IPPP/16/53 \\
\today \\
\end{flushright}

\vspace*{0.5cm}
\begin{center}
{\Large \bf  Jet activity as a probe of high--mass resonance \\
\vspace*{0.4cm}
production}\\

\vspace*{1cm}

L. A. Harland-Lang$^a$, V. A. Khoze$^{b,c}$, M. G. Ryskin$^c$ and M. Spannowsky$^b$

\vspace*{0.5cm}
$^a$ Department of Physics and Astronomy, University College London, WC1E 6BT, UK \\           
$^b$ Institute for Particle Physics Phenomenology, Durham University, DH1 3LE, UK    \\
$^c$
 Petersburg Nuclear Physics Institute, NRC Kurchatov Institute, 
Gatchina, St.~Petersburg, 188300, Russia \\
\end{center}

\begin{abstract}
\noindent We explore the method of using the measured jet activity associated with a high mass resonance state to determine the corresponding production modes. To demonstrate the potential of the approach, we consider the case of a resonance of mass $M_R$ decaying to a diphoton final state. We perform a Monte Carlo study, considering three mass points $M_R=0.75,\,1.5\,,2.5$ TeV, and show that  the $\gamma\gamma$, $WW$, $gg$ and light and heavy $q\overline{q}$ initiated cases lead to distinct predictions for the jet multiplicity distributions. As an example, we apply this result to the ATLAS search for resonances in diphoton events, using the 2015 data set of $3.2\,{\rm fb}^{-1}$ at $\sqrt{s}=13$ TeV. Taking the spin--0 selection, we demonstrate that a dominantly $gg$--initiated signal hypothesis is mildly disfavoured, while the $\gamma\gamma$ and light quark cases give good descriptions within the limited statistics, and a dominantly $WW$--initiated hypothesis is found to be in strong tension with the data. We also comment on the $b\overline{b}$ initial state, which can already be constrained by the measured $b$--jet multiplicity. Finally, we present expected exclusion limits with integrated luminosity, and demonstrate that with just a few 10's of ${\rm fb}^{-1}$ we can expect to constrain the production modes of such a resonance.
\end{abstract}
\vspace*{0.5cm}

\section{Introduction}

Both the ATLAS and CMS collaborations reported the observation of an excess of events in the diphoton mass distribution around 750 GeV~\cite{Aaboud:2016tru,Khachatryan:2016hje} in roughly $3\,{\rm fb}^{-1}$ of data recorded in 2015 at $\sqrt{s}=13$ TeV, which was found to be compatible with the data collected at 8 TeV. The possibility that this corresponded to a new resonance state generated a great deal of theoretical interest, and a wide range of BSM models describing it were  proposed, see for example~\cite{Strumia:2016wys} for a review and further references. While no significant excess was observed in the larger data set collected in 2016~\cite{ATLAS:2016eeo,Khachatryan:2016yec}, it is nonetheless interesting to consider how a potential resonance of this type, which might be present at higher mass and hence be observable at the LHC or a future collider, can be produced.

One interesting, and in some sense natural possibility, for an excess that is seen in the $\gamma\gamma$ decay channel, is that the resonance may couple dominantly to photons, with the coupling to gluons and other coloured particles being either suppressed or absent entirely. This has been discussed in~\cite{Fichet:2015vvy,Csaki:2015vek,Csaki:2016raa,Harland-Lang:2016qjy,Abel:2016pyc,Backovic:2016gsf,Mandal:2016dsx,Anchordoqui:2016rve,Bernon:2016dow,Dobrescu:2016owc}. More generally, we may expect significant couplings to quarks and gluons to be present. A method to distinguish between these different production modes, discussed in for instance~\cite{Gao:2015igz,Csaki:2016raa,Harland-Lang:2016qjy,Ebert:2016idf,Dalchenko:2016dfa}, is to measure the multiplicity of  jets produced in addition to the resonance. In this case we may naturally anticipate that if the resonance is dominantly produced in $\gamma\gamma$ collisions, then the average jet multiplicity will be lower compared to the $q\overline{q}$ and $gg$ cases. While additional jets in this case are \textit{not} parametrically suppressed by $\sim\alpha/\alpha_s$ -- no additional $O(\alpha)$ suppression is introduced by requiring that the quark produced in the initial--state $q\to q\gamma$ splitting leads to a visible jet in the final--state -- nonetheless some suppression is present due to the smaller size of $\alpha$ and lower photon branching probability. It is in addition well known that the particle multiplicity associated with an initial--state gluon is higher compared to the quark case, and so again some difference in jet activity can be expected here. For the vector boson fusion (VBF) $WW$--initiated channel the resonance is generally produced in association with at least two additional jets, due to the relatively high $p_\perp$ recoiling quarks in the final state.

With this in mind, in this paper we we will consider the case of a scalar resonance $R$ for three mass points $M_R=0.75,\,1.5\,,2.5$ TeV, to demonstrate the viability of the method, although this can be readily extended to other spin--parities. We will present a detailed analysis of the expected jet multiplicities corresponding to the different production scenarios, focussing on the $gg$, light $q\overline{q}$ and heavy $b\overline{b}$, $\gamma\gamma$ and $WW$ cases. We will show that the above expectations are indeed born out by a more precise MC analysis, which accounts for the decay of the resonance and full experimental acceptances and jet selection.

To demonstrate how such an approach may be applied to data, we will then compare to the 2015 ATLAS~\cite{Aaboud:2016tru} measurement of the jet multiplicity in the spin--0 event selection sample, and demonstrate that this limited statistics data already show some mild tension with a dominantly $gg$--initiated scenario, while the light quark and in particular $\gamma\gamma$--initiated scenarios give good descriptions. The continuum background--only hypothesis also gives a somewhat worse description than these latter cases, and the $WW$ hypothesis is found to give a particularly poor description, with a dominantly $WW$--initiated production mechanism in strong tension with the data; a similar conclusion will hold for the $ZZ$, and to a lesser extent, $Z\gamma$--initiated mechanisms. In the $b\overline{b}$--initiated case there is expected to be a sizeable fraction of $b$--jets observed in the final state, rendering such a possibility relatively easy to confirm or constrain, even for this limited data set; indeed the ATLAS measurement of the $b$--jet fraction in the signal region disfavours any sizeable $bb$--induced production mode. We also compare in an appendix to the 2015 ATLAS data collected with a spin--2 event selection, but still assuming a scalar resonance signal. While the total number of observed events in the signal region is larger, the $S/B$ ratio is lower, and we find that all hypotheses give acceptable descriptions of the data.

Although no excess at $750$ GeV was observed in the updated 2016 data set, this nonetheless demonstrates the discriminating power of this observable. For this reason we will also present expected exclusion limits on different production scenarios with integrated luminosity, and will show that with just a few $10$'s of ${\rm fb}^{-1}$ it is possible to place strong constraints on the $gg$ production mechanism in favour of the $\gamma\gamma$/$q\overline{q}$ cases, and vice versa. It is more challenging to distinguish between the $u\overline{u}$ and $\gamma\gamma$ modes via this method, although still possible with enough data. 

The outline of this paper is as follows. In Section~\ref{sec:analysis} we present the details of our analysis and the simplified production model we consider. In Section~\ref{sec:jetmulti} we present results for the predicted jet multiplicities corresponding to the different production scenarios. In Section~\ref{sec:ATLAS} we compare to the ATLAS jet multiplicity measurement, and comment on the implications for the various production modes. In Section~\ref{sec:exc} we present the expected exclusion limits on different production scenarios as a function of the integrated luminosity collected at the LHC. Finally, in Section~\ref{sec:conc} we conclude. In Appendix~\ref{sec:ATLAS2} we compare to the ATLAS spin--2 event selection, and show that all initial state hypotheses provide acceptable descriptions.

\section{Production model and analysis}\label{sec:analysis}

To model the production of a scalar resonance $X\to \gamma\gamma$ via $\gamma\gamma$, $gg$ and $q\overline{q}$ initiated production  we use an effective theory approach, with corresponding Lagrangian terms
\begin{align}\nonumber
\mathcal{L}^{gg}&=g_{g} \,G^{\mu\nu}G_{\mu\nu}\,R\;,\\ \nonumber
\mathcal{L}^{\gamma\gamma}&=g_{\gamma} \,F^{\mu\nu}\,F_{\mu\nu}R\;,\\ \nonumber
\mathcal{L}^{q\overline{q}}&=g_q\, q\overline{q}R\;,\\ 
\mathcal{L}^{WW}&=g_W\, W^{\mu\nu}\,W_{\mu\nu}R\;,
\end{align}
where we make no assumptions about the size of the  couplings $g$. We consider resonance masses of $M_R=0.75,\,1.5,\,2.5$ TeV, with a uniform width of $\Gamma_{\rm tot}=45$ GeV, although the results which follow are largely independent of the size of the width.

Parton--level events are then generated at LO with up to 2 additional partons in the final--state using \texttt{MadGraph 5}~\cite{Alwall:2014hca}, which are MLM merged at scale $Q_{\rm cut}=125,\,250,\,350$ GeV for $M_R=0.75,\,1.5,\,2.5$ TeV, respectively, to parton shower generated with \texttt{Pythia 8}~\cite{Sjostrand:2006za,Sjostrand:2007gs}, including hadronization and multiple parton interactions. Events for the Standard Model continuum $\gamma\gamma$ process, which proceeds dominantly via $q\overline{q}\to \gamma\gamma$, are generated in the same way. For the $WW$ initial state only the Born--level process, which generally leads to two additional quark jets in the final state, is generated and passed to \texttt{Pythia} for parton shower and hadronization, with no matching required. For the initial--state photon PDFs we apply the approach described in~\cite{Harland-Lang:2016apc,Harland-Lang:2016qjy}, with the MMHT14 LO~\cite{Harland-Lang:2014zoa} PDF set used for all other partons, however the normalized distributions we present below are relatively insensitive to these choices. All the analyses performed below make use of \texttt{Rivet}~\cite{Buckley:2010ar}. 

For concreteness,  we apply the event selection:

\begin{itemize}
\item Two reconstructed photons, satisfying $p_\perp>0.4(0.3) M_{\gamma\gamma}$ for the leading (subleading) photon and $|\eta_\gamma|<2.7$.
\item The diphoton invariant mass lies in the range $M_R-25 \,{\rm GeV} <M_{\gamma\gamma}<M_R+25$ GeV.
\item An isolation requirement for all particles in a cone $\Delta R=0.4$ around the photons direction $E_\perp^{\rm iso}<0.05 E_\perp^\gamma + 6$ GeV.
\end{itemize}
This is guided by the ATLAS~\cite{Aaboud:2016tru} selection for the spin--0 resonance sample, however we note that the results which follow are largely insensitive to these precise choices, for example, if a looser set of cuts on the final--state photons is imposed. We reconstruct jets with the anti--$k_t$ algorithm~\cite{Cacciari:2008gp} with distance parameter $R=0.6$, $p_\perp^j>25$ GeV and $|\eta^j|< 4.4$. We choose this somewhat larger value of $R$ compared to that taken in~\cite{Aaboud:2016tru}  as this allows a greater discrimination between the different production modes, while still being relatively insensitive to the influence of the underlying event generated with \texttt{Pythia} for the corresponding minimum jet $p_\perp$.

\section{Jet multiplicities}\label{sec:jetmulti}

\begin{figure}[h]
\begin{center}
\includegraphics[scale=0.65]{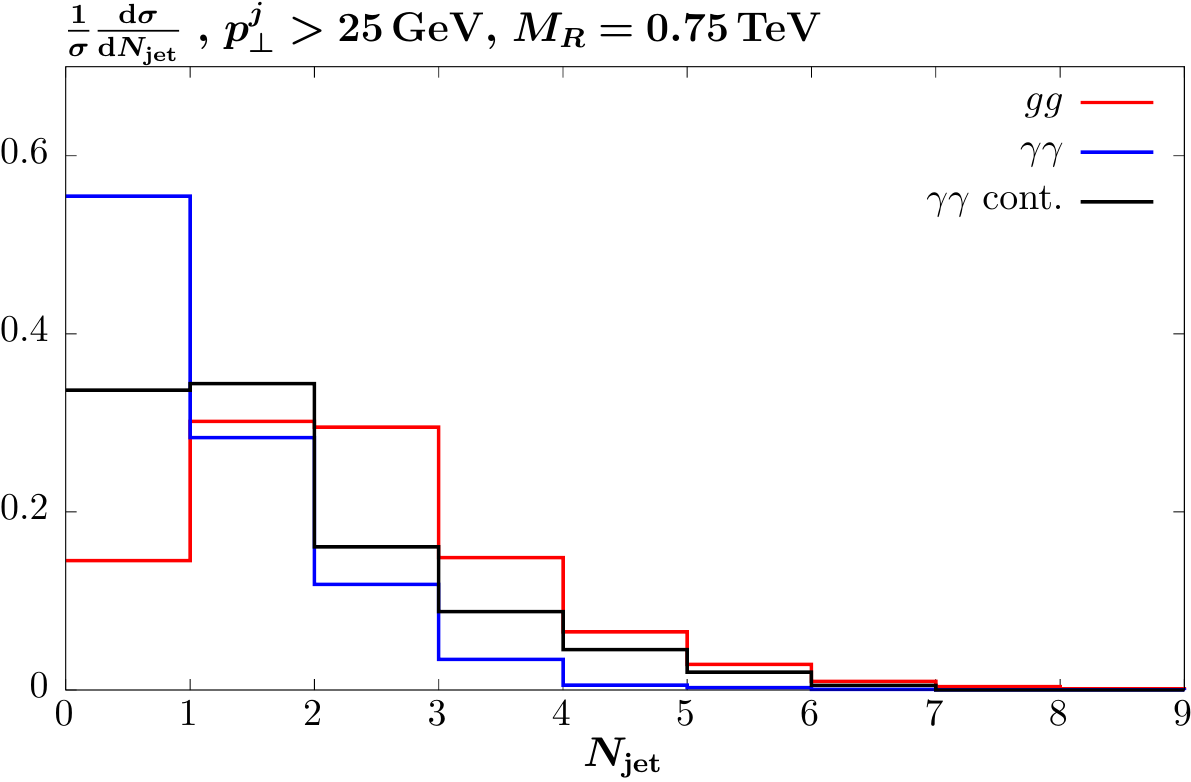}
\includegraphics[scale=0.65]{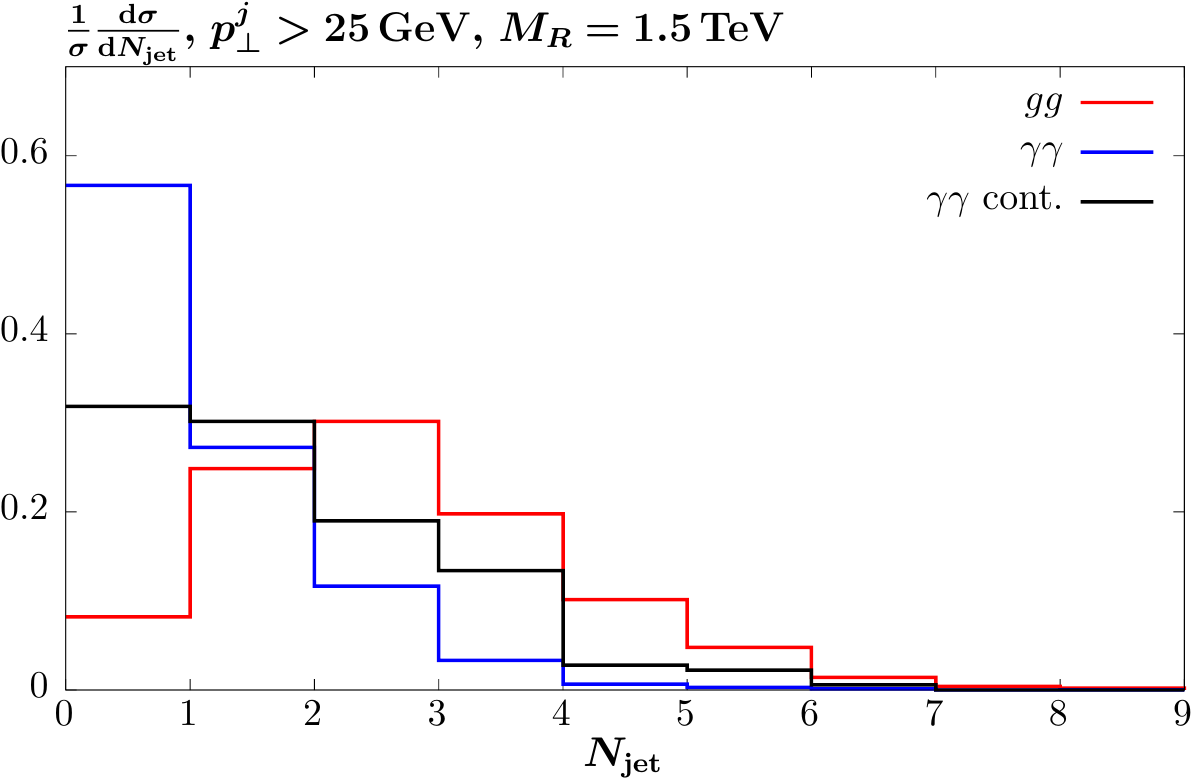}
\includegraphics[scale=0.65]{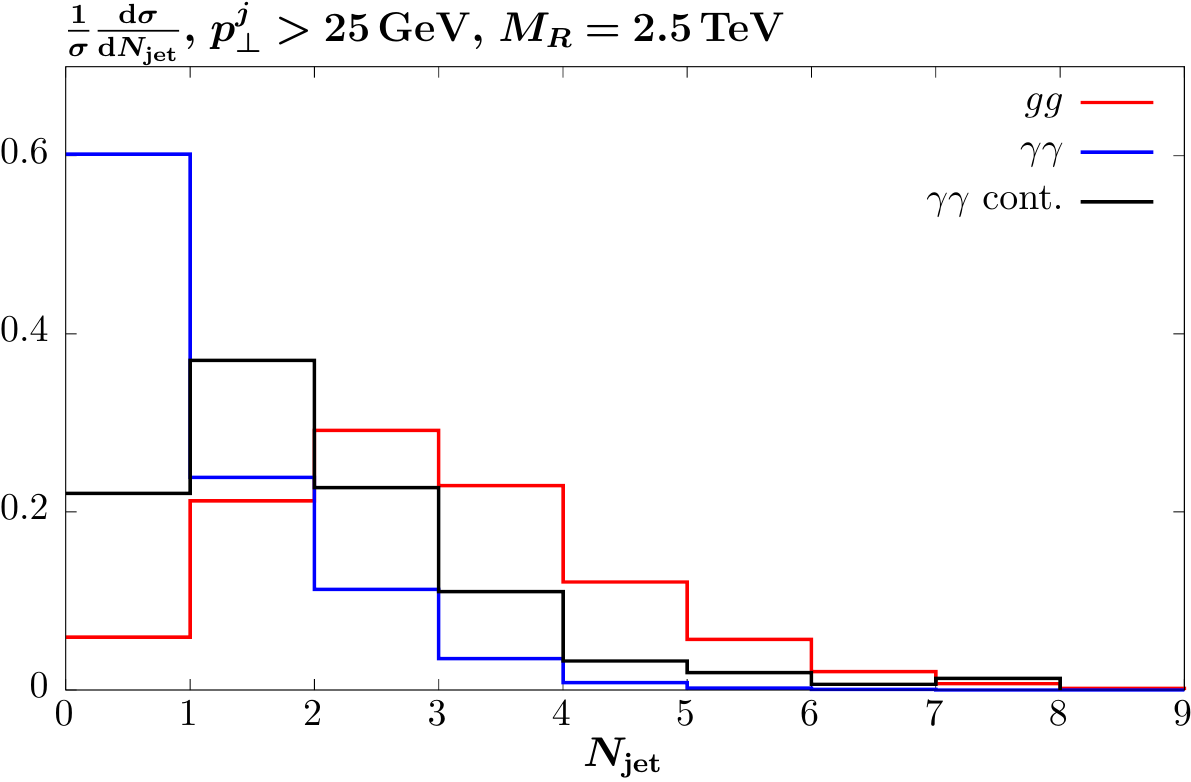}
\caption{Exclusive jet multiplicities for minimum jet $p_\perp>25$ GeV, for the $gg$ and $\gamma\gamma$ initial--state resonance production processes, and the SM continuum $\gamma\gamma$ production process. Results shown for resonance masses $M_R=0.75,\,1.5,\,2.5$ TeV.}
\label{fig:jetmulti}
\end{center}
\end{figure} 

\begin{figure}[h]
\begin{center}
\includegraphics[scale=0.65]{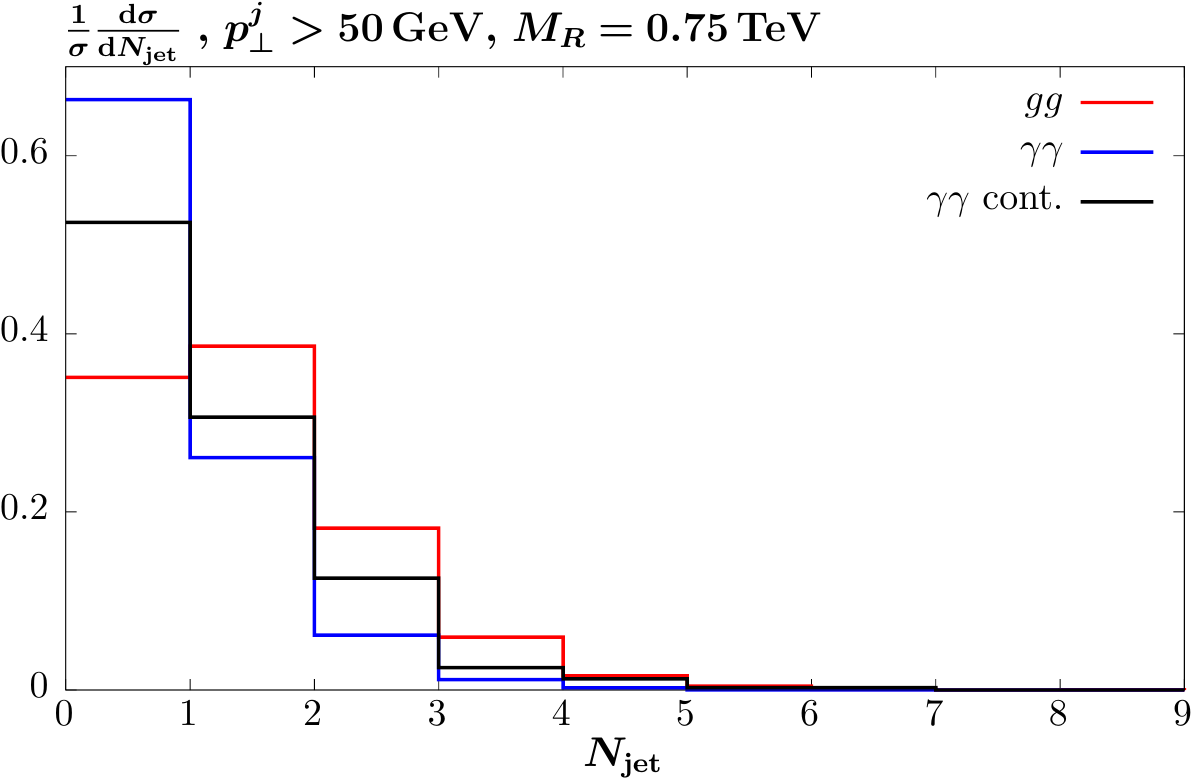}
\includegraphics[scale=0.65]{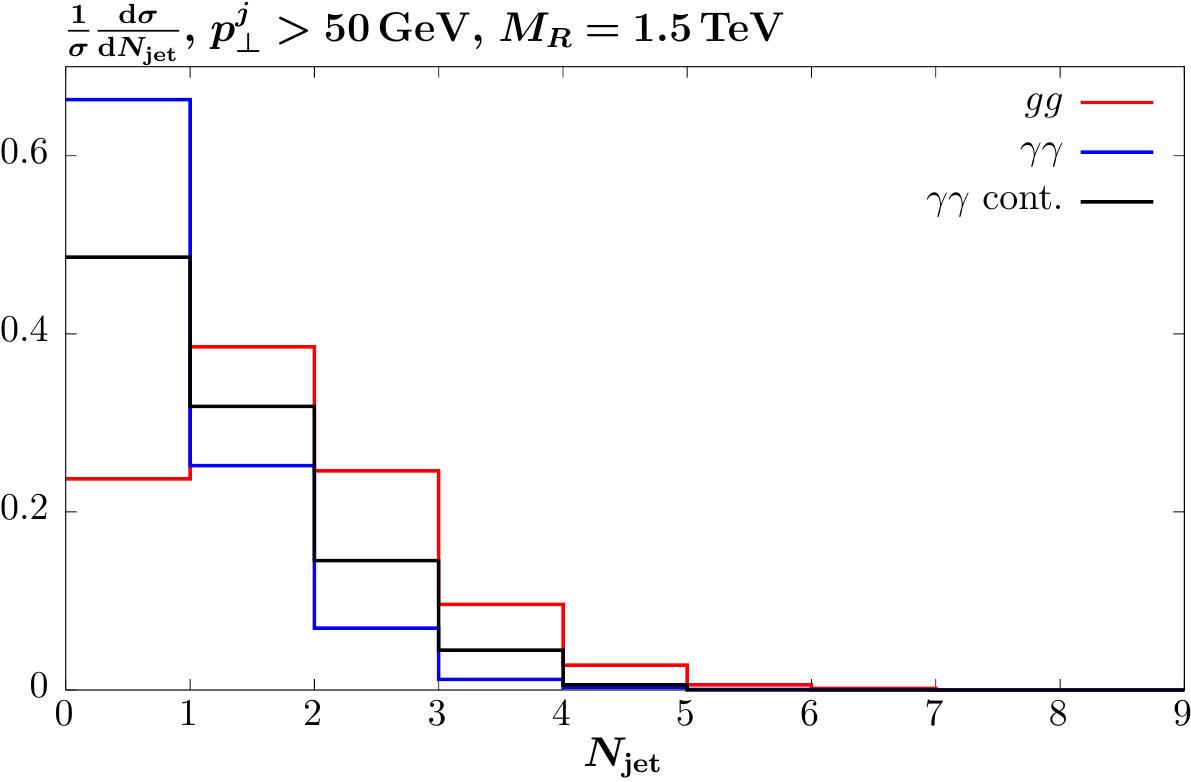}
\includegraphics[scale=0.65]{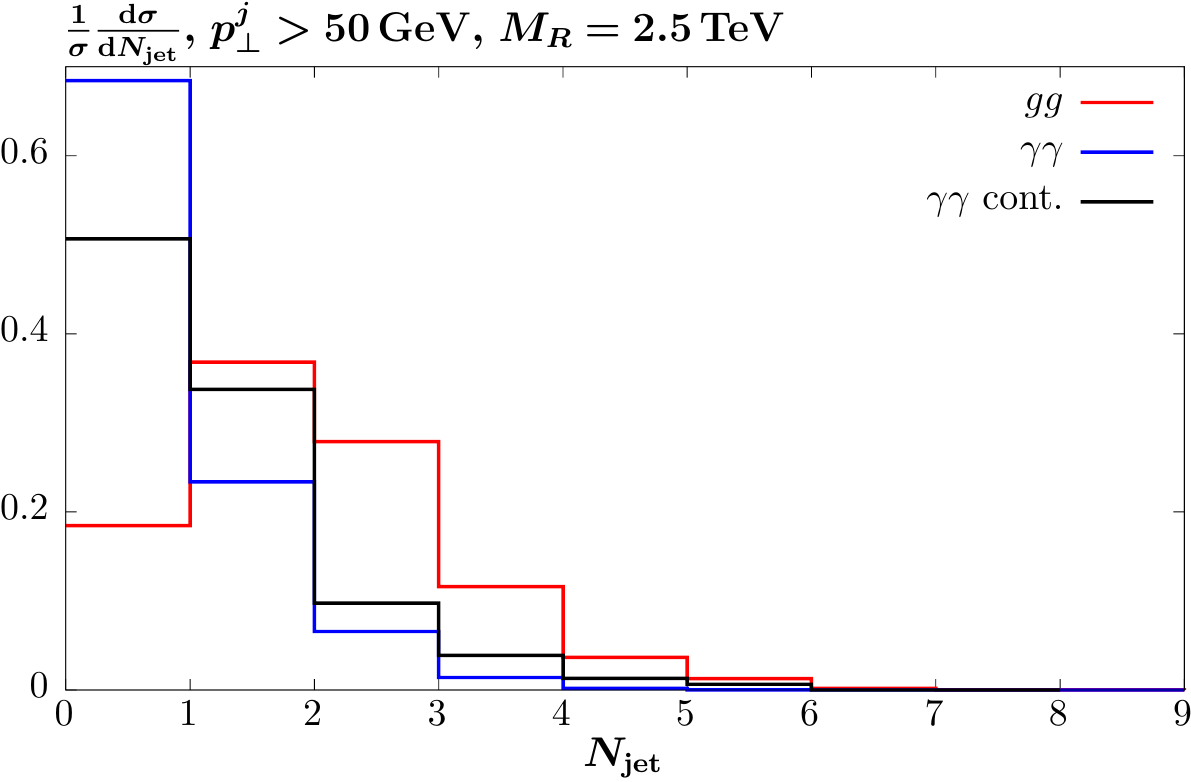}
\caption{Exclusive jet multiplicities for minimum jet $p_\perp >50$ GeV, for the $gg$ and $\gamma\gamma$ initial--state resonance production processes, and the SM continuum $\gamma\gamma$ production process. Results shown for resonance masses $M_R=0.75,\,1.5,\,2.5$ TeV.}
\label{fig:jetmultipt50}
\end{center}
\end{figure} 

Using the event selection and model choice outlined in the previous section we can then evaluate the cross section for resonance production via the $\gamma\gamma$, $gg$ and $q\overline{q}$ initial--states. In the latter case we will consider both $u\overline{u}$ and $b\overline{b}$, since as we will see there is a non--negligible difference in the expected distributions for light and heavy quark mediated processes. The expected distributions for the other light $d\overline{d}$ and $s\overline{s}$ quark initiated modes are very similar to the $u\overline{u}$, and thus while we will for concreteness refer in what follows to the `$u\overline{u}$' production mode, this can be considered as the prediction for any light quark--induced process. 

In addition, as discussed in~\cite{Dalchenko:2016dfa} this method may also be applied to the case of, for example, $WW$--initiated VBF. We will consider this below, but it is worth pointing out that, as discussed in~\cite{Harland-Lang:2016qjy}, due to the relatively large mass of the exchanged $t$--channel $W$ bosons the transverse momentum of the final--state $\gamma\gamma$ system may also be a sensitive observable. In particular within the simple approach of that paper we expect that only roughly $\sim 20$\% of the cross section has $p_\perp^{\gamma\gamma} < 60$ GeV. The same effect will also be similarly present in the $ZZ$ and $Z\gamma$ channels. 

\begin{figure}[h]
\begin{center}
\includegraphics[scale=0.65]{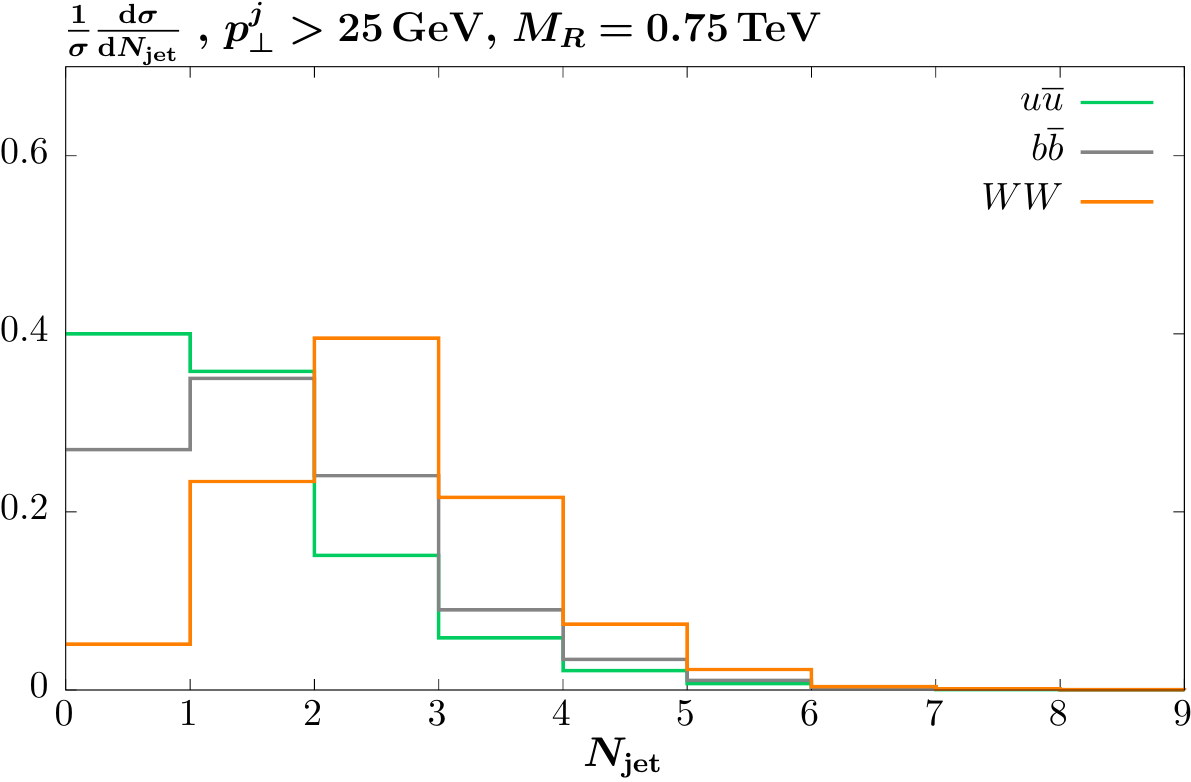}
\includegraphics[scale=0.65]{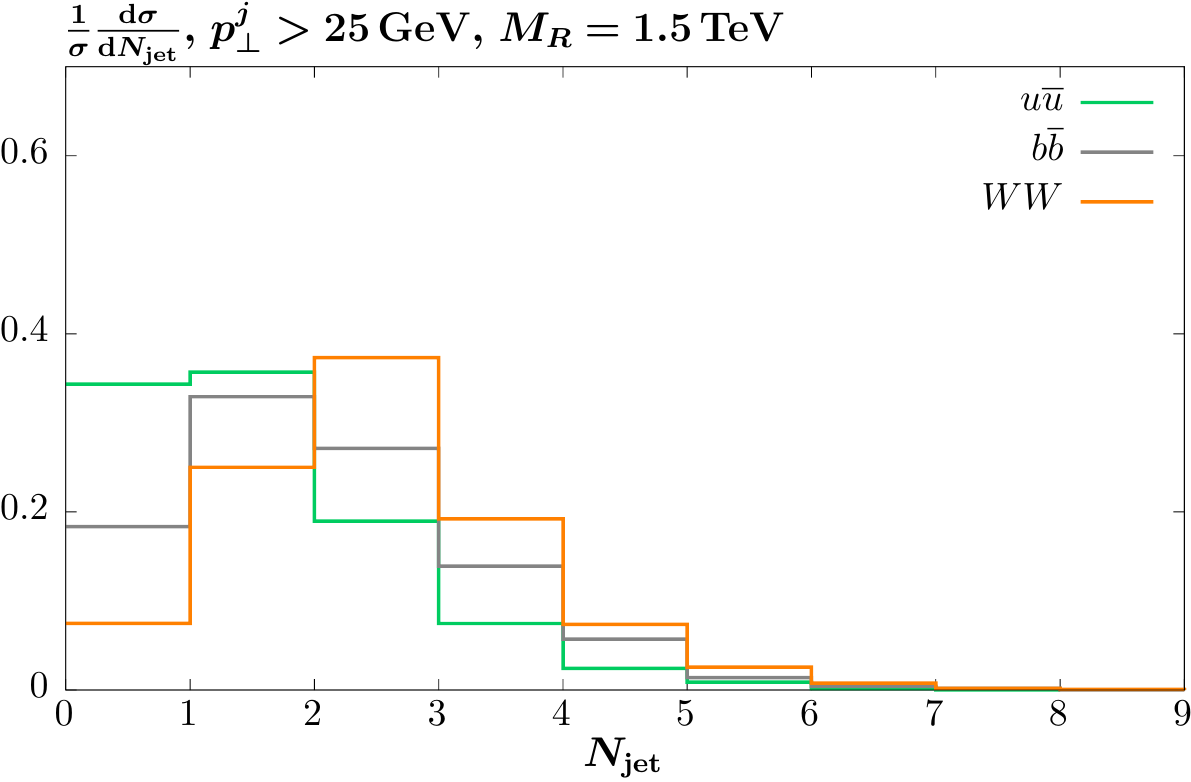}
\includegraphics[scale=0.65]{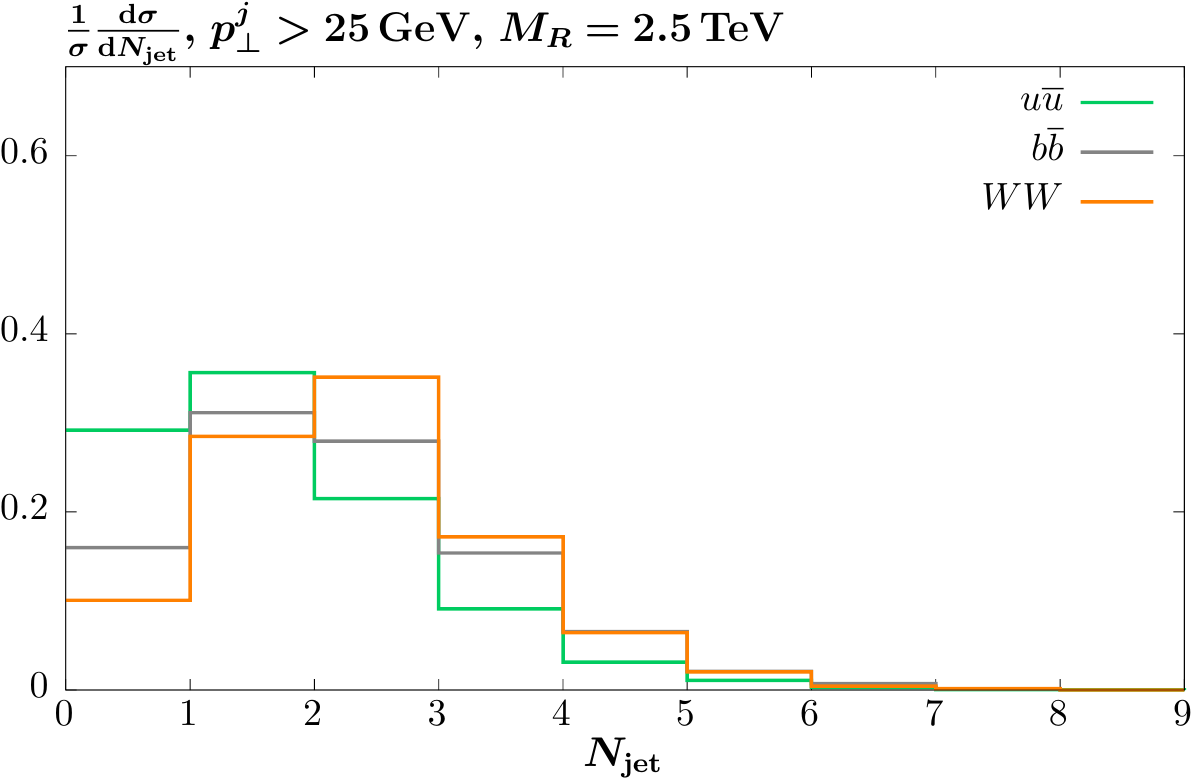}
\caption{Exclusive jet multiplicities for minimum jet $p_\perp>$ 25 GeV, for 
the $u\overline{u}$, $b\overline{b}$ and $WW$ initial--state resonance production processes. Results shown for resonance masses $M_R=0.75,\,1.5,\,2.5$ TeV.}
\label{fig:jetmultiq}
\end{center}
\end{figure} 

\begin{figure}[h]
\begin{center}
\includegraphics[scale=0.65]{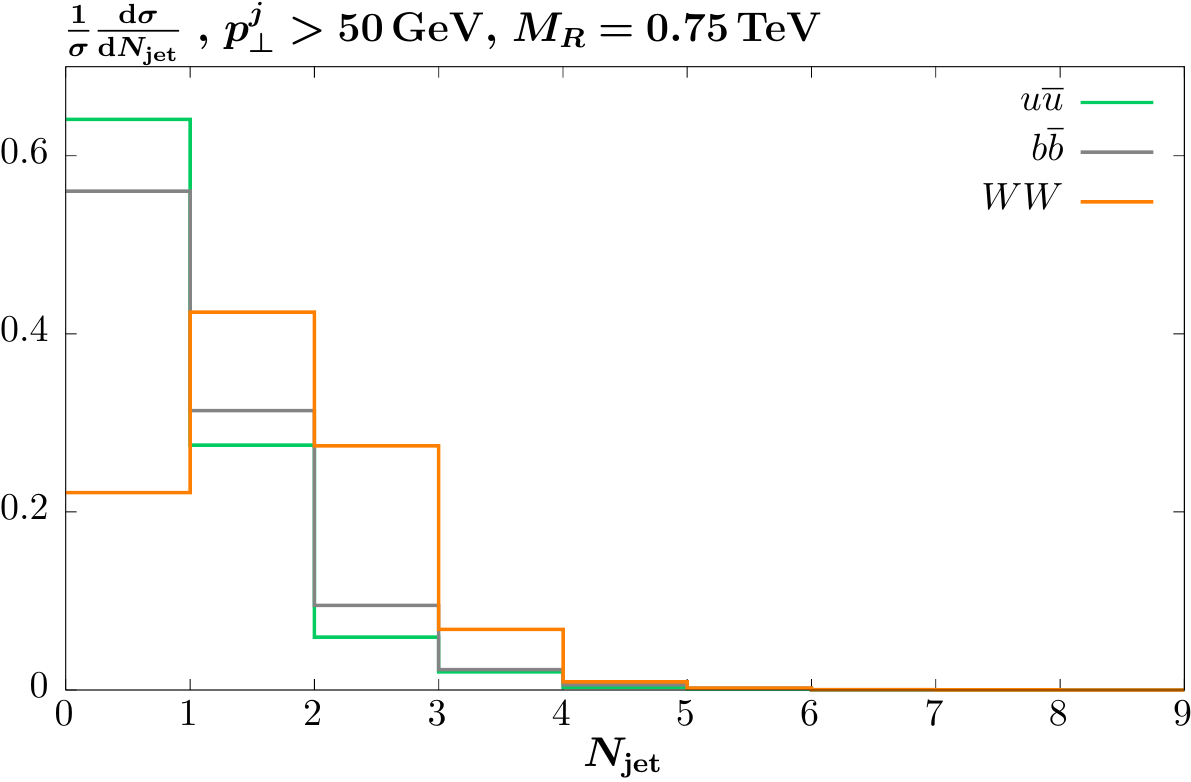}
\includegraphics[scale=0.65]{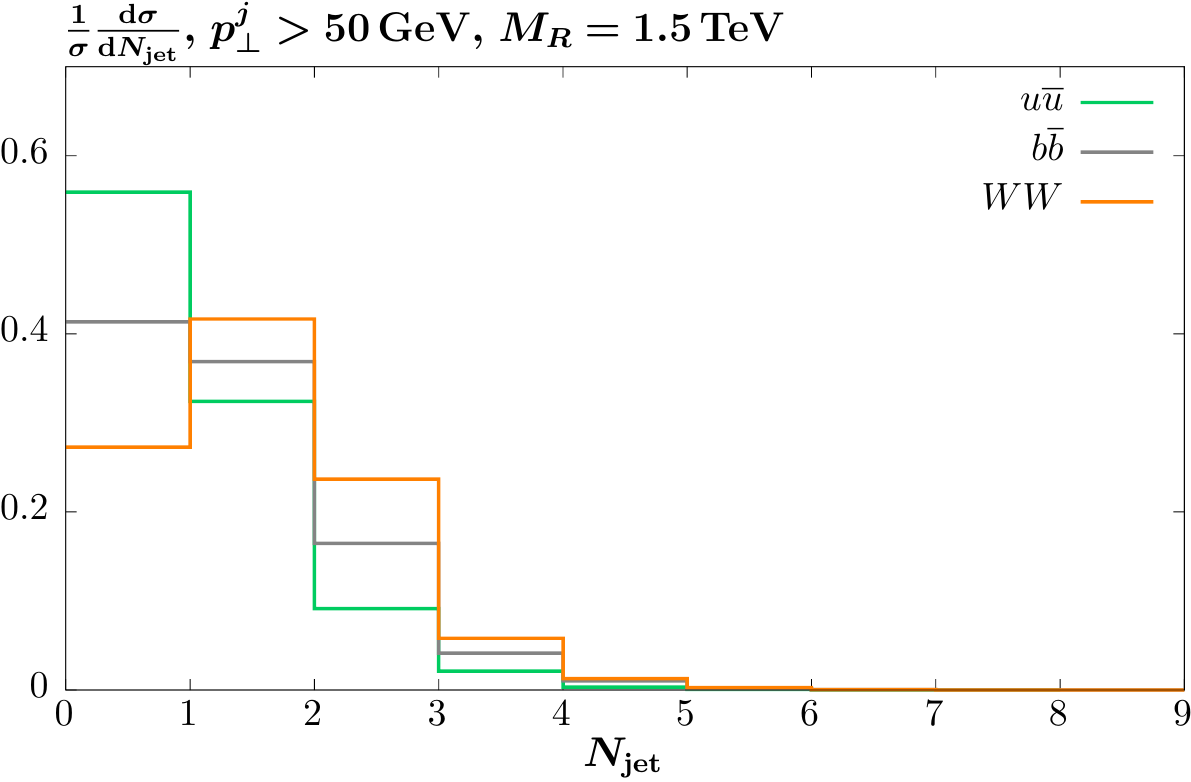}
\includegraphics[scale=0.65]{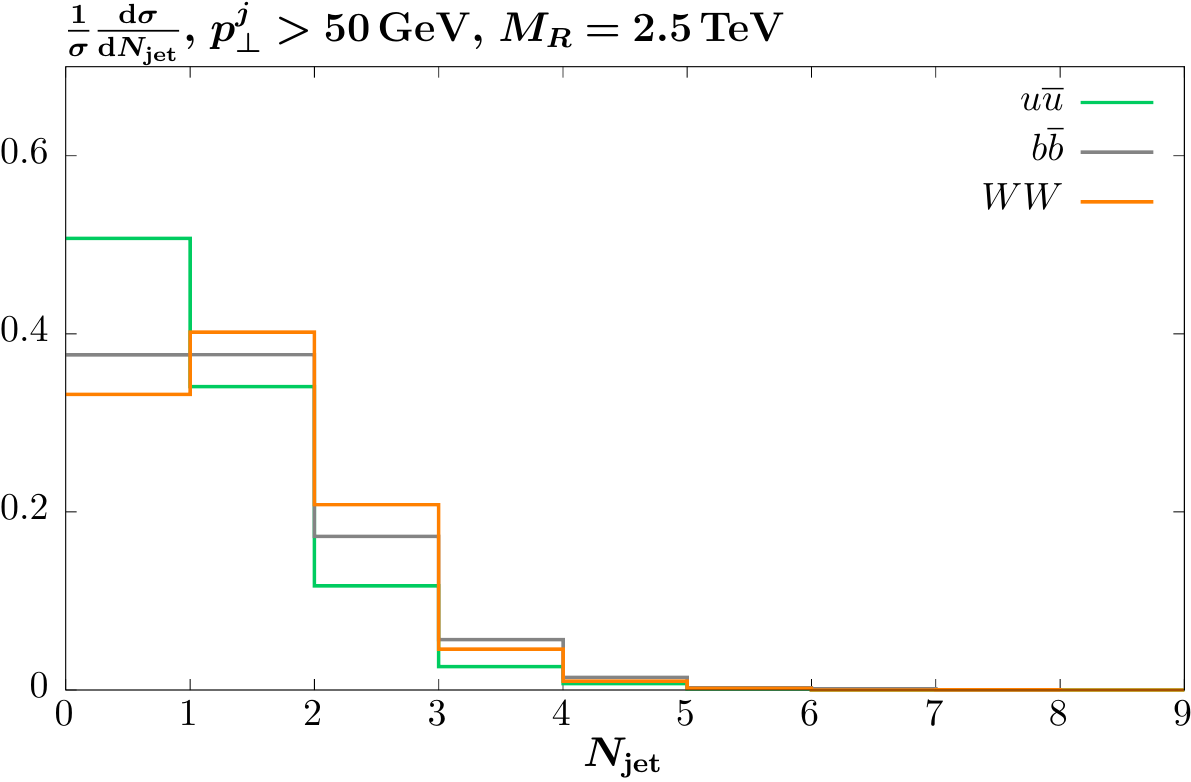}
\caption{Exclusive jet multiplicities for minimum jet $p_\perp>$ 50 GeV, for the $u\overline{u}$, $b\overline{b}$ and $WW$ initial--state resonance production processes. Results shown for resonance masses $M_R=0.75,\,1.5,\,2.5$ TeV.}
\label{fig:jetmultiqpt50}
\end{center}
\end{figure} 

In Figs.~\ref{fig:jetmulti} and~\ref{fig:jetmultipt50} we show the cross section for resonance production accompanied by $N_{\rm jet}$ jets, for $p_\perp^j>$ 25 (50) GeV, respectively. The results are presented with the fixed merging scales described in the previous section, but we have checked that when varying these between a wider range the results presented below do not vary by more than $\sim 10\%$, and usually much lower; this can be considered as an estimate of the theoretical uncertainty in our approach, and all results should be interpreted with this  in mind.  We can see that the difference between the $gg$ and $\gamma\gamma$ initial--states for the lower $p_\perp^j>$ 25 GeV cut is dramatic, with over $50\%$ of the $\gamma\gamma$ initiated events having no jets passing the selection, while for the $gg$--initiated process this is below $20\%$. In addition, the overall shape of the jet multiplicity distribution is correspondingly different, with the $\gamma\gamma$ distribution peaking at 0 jets, while the $gg$  exhibits a peak at $N_{\rm jet}=1-2$. At higher $p_\perp^j>$ 50 GeV the difference is less dramatic, although the population of the 0 jet bin is still almost a factor of $2$ higher in the $\gamma\gamma$ case. The cause of this remarkably different behaviour is discussed in detail in~\cite{Harland-Lang:2016qjy}, and is generated by the simple fact that the initial--state coloured gluons exhibit a strong preference to radiate further, so that there is a strong Sudakov suppression in the cross section for no further emission within the corresponding acceptance; the initial--state photons, on the other hand, exhibit a much weaker preference to radiate further, and while for example the parent quark in the $q\to q\gamma$ splitting may lead to an observable jet in the final--state, the effect of the jet veto is much less significant. A similar trend is seen in all three mass points, although in general the jet multiplicity is observed to increase with increasing mass for the $gg$ and continuum cases, as expected from the increasing phase space for bremstrahlung gluon emission. Interestingly, for the $\gamma\gamma$ initial state the multiplicity in fact decreases as $M_R$ is increases. A possible explanation for this effect may be the fact that to very good approximation the only observed jets here are due to the final $q\to q\gamma$ splitting before the hard process; the $N_j \geq 3$ population is extremely low. As $M_R$ increases the average momentum fraction $x$ carried by the parent quark, and hence the produced quark jet, increases. This will lead to the jet being produced on average at higher rapidity, making it more like to fail the $|\eta_j|<4.4$ requirement. As a result of these trends, for a $M_R=2.5$ GeV resonance the $\gamma\gamma$ to $gg$ ratio in the 0 jet bin is particularly large, and the discriminating power of this approach improves with increasing resonance mass.

\begin{figure}[h]
\begin{center}
\includegraphics[scale=0.65]{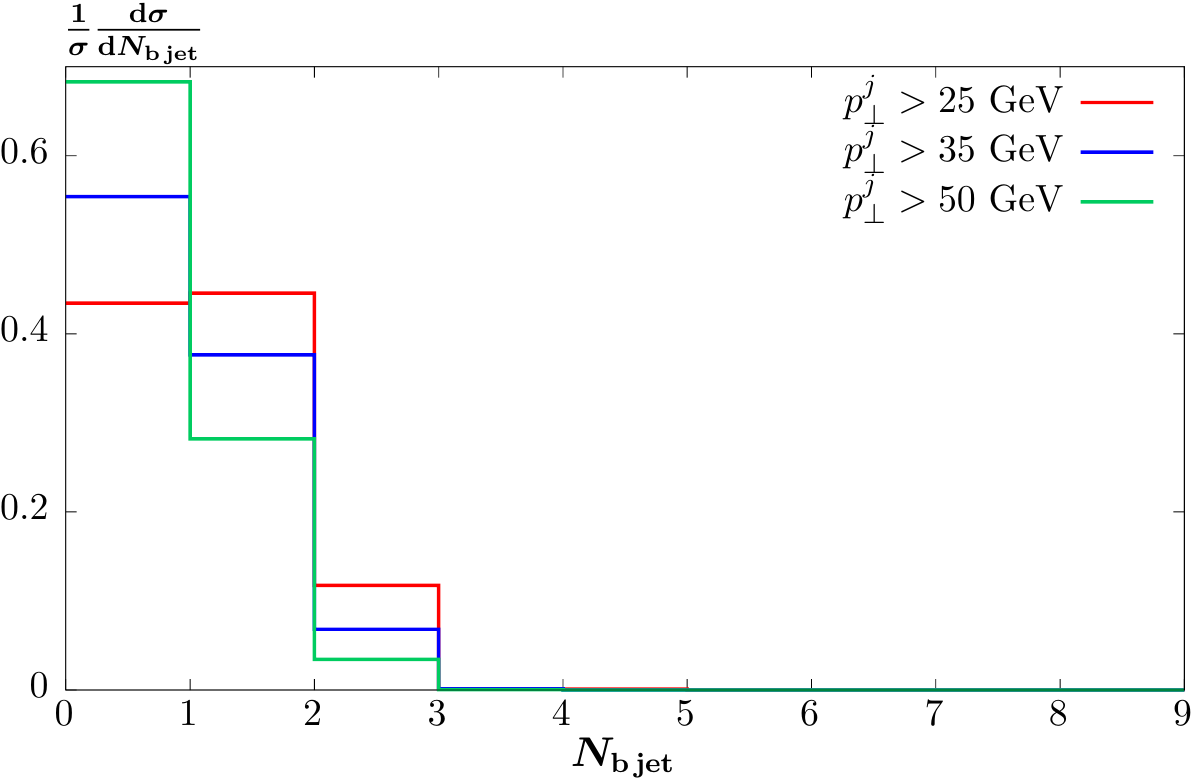}
\caption{Exclusive $b$ jet multiplicities for different minimum jet $p_\perp$, for the $b\overline{b}\to R$ production process and for resonance mass  $M_R=0.75$ TeV.}
\label{fig:bjetmulti}
\end{center}
\end{figure} 

In Figs.~\ref{fig:jetmultiq} and~\ref{fig:jetmultiqpt50} we show the predicted distribution for the $WW$--initiated VBF channel for $p_\perp^j>$ 25 (50) GeV, respectively: here, the resonance is produced in association with two outgoing quarks recoiling against the exchanged $W$ bosons, and which carry on average a relatively high transverse momentum $p_\perp \sim M_W$. These therefore tend to produce observable jets in the final state, so that the predicted jet multiplicity is very high, larger still than in the $gg$ case, and strongly peaking at $N_{\rm jet}=2$. As in the $\gamma\gamma$ case, as the resonance mass is increased the average multiplicity is seen to decrease, with the 0 and 1 jet bins become more populated.

\begin{figure}[h]
\begin{center}
\includegraphics[scale=0.65]{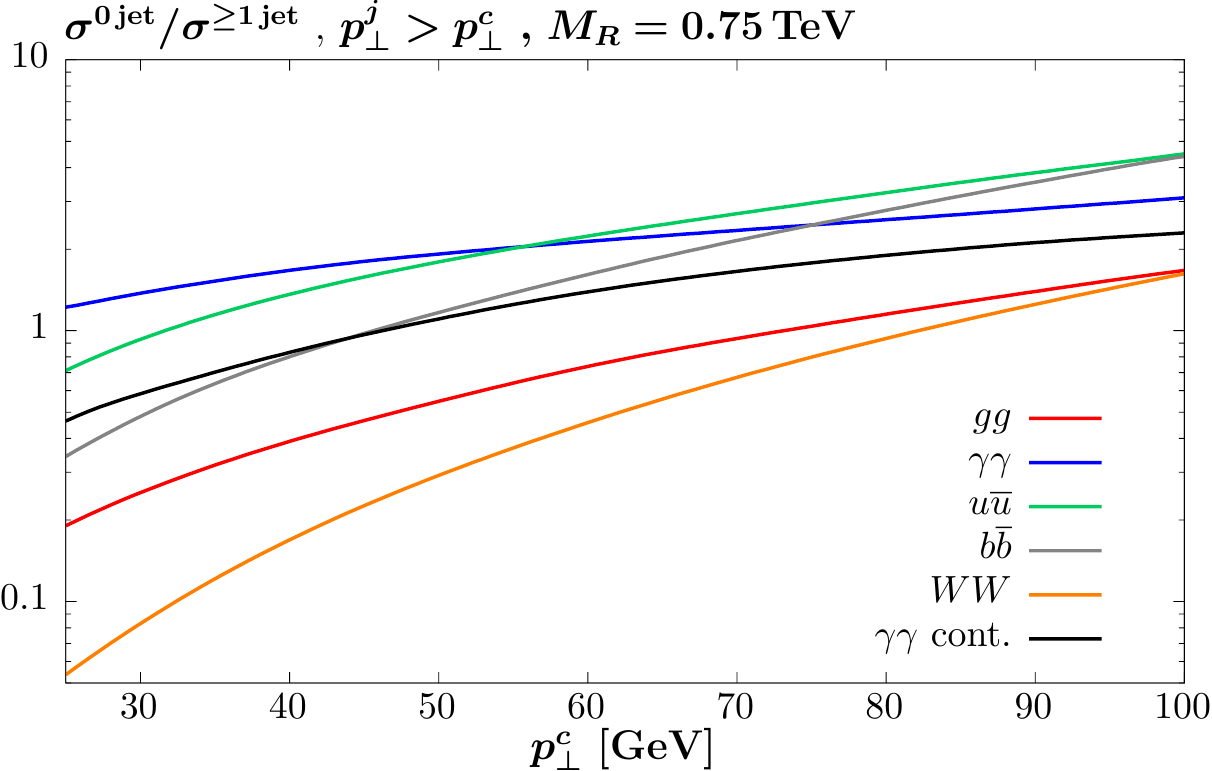}
\includegraphics[scale=0.65]{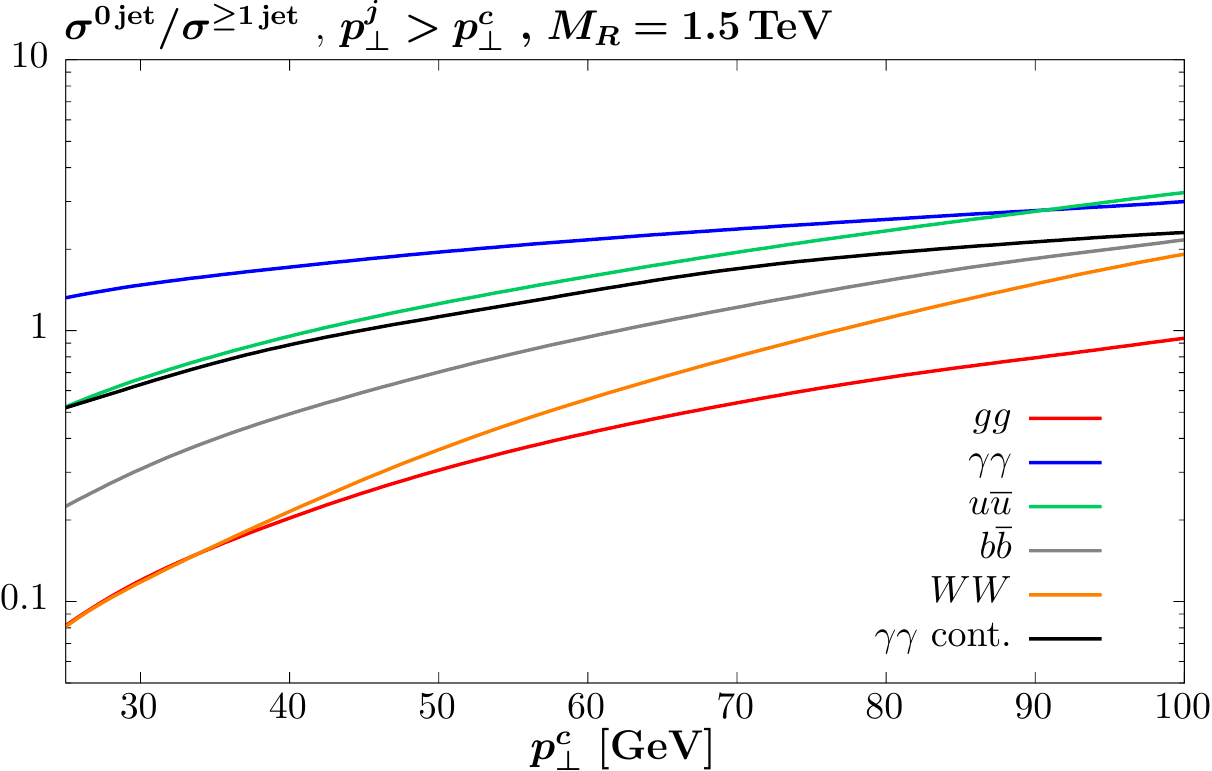}
\includegraphics[scale=0.65]{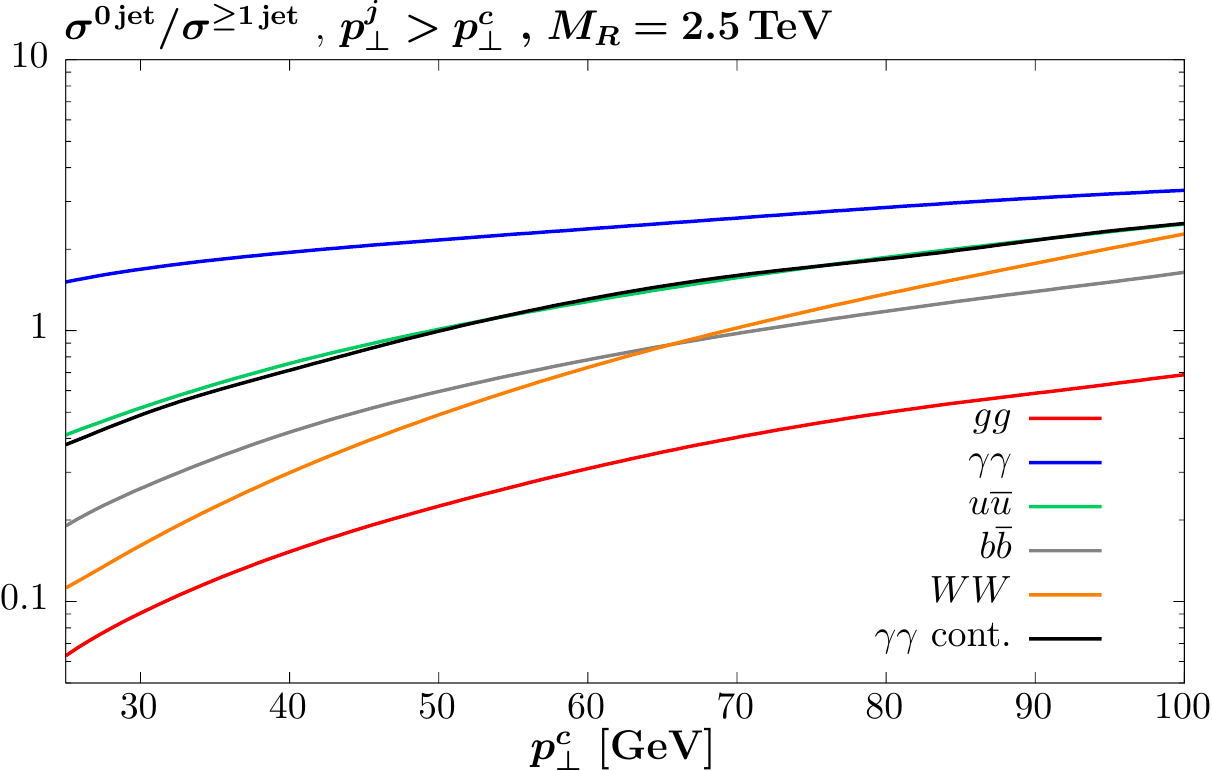}
\caption{Ratio of 0 jet to $\geq$ 1 jet cross sections as a function of the minimum jet $p_\perp$ for a range of initial--state resonance production processes, and for the continuum $\gamma\gamma$ background. Results shown for resonance masses $M_R=0.75,\,1.5,\,2.5$ TeV.}
\label{fig:jetptveto}
\end{center}
\end{figure} 

\begin{figure}[h]
\begin{center}
\includegraphics[scale=0.65]{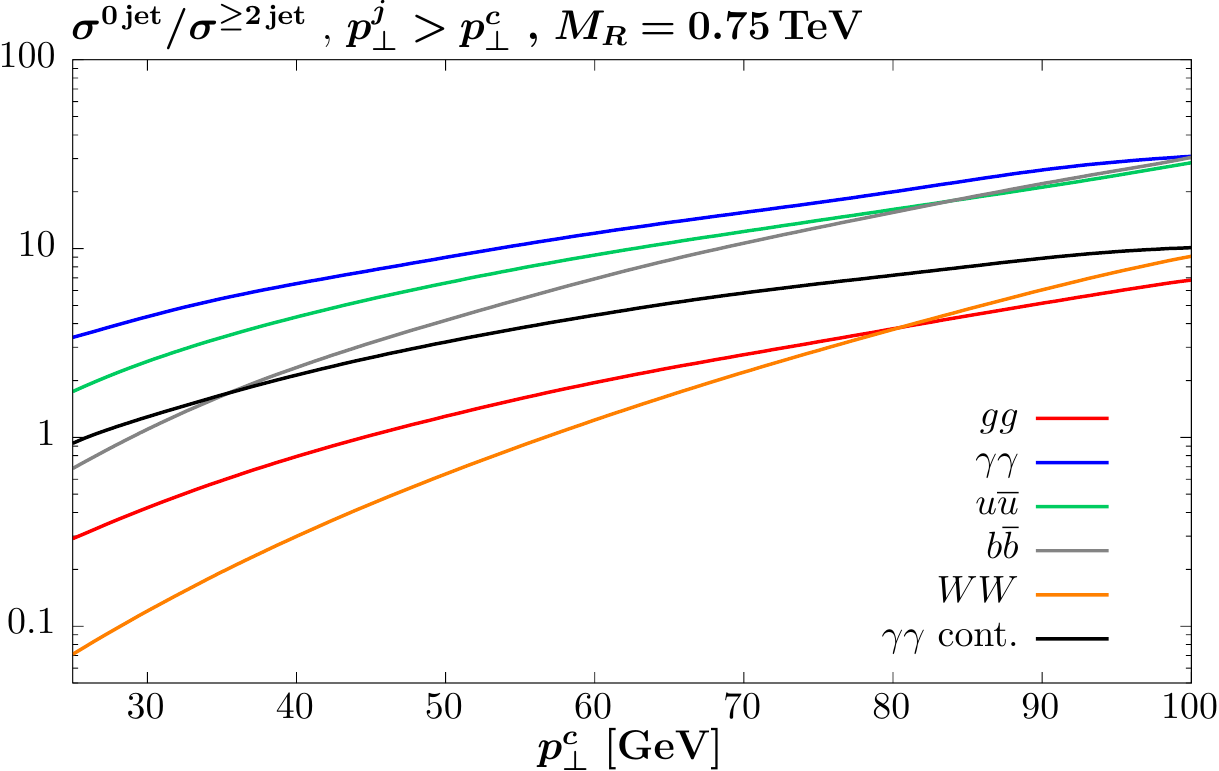}
\includegraphics[scale=0.65]{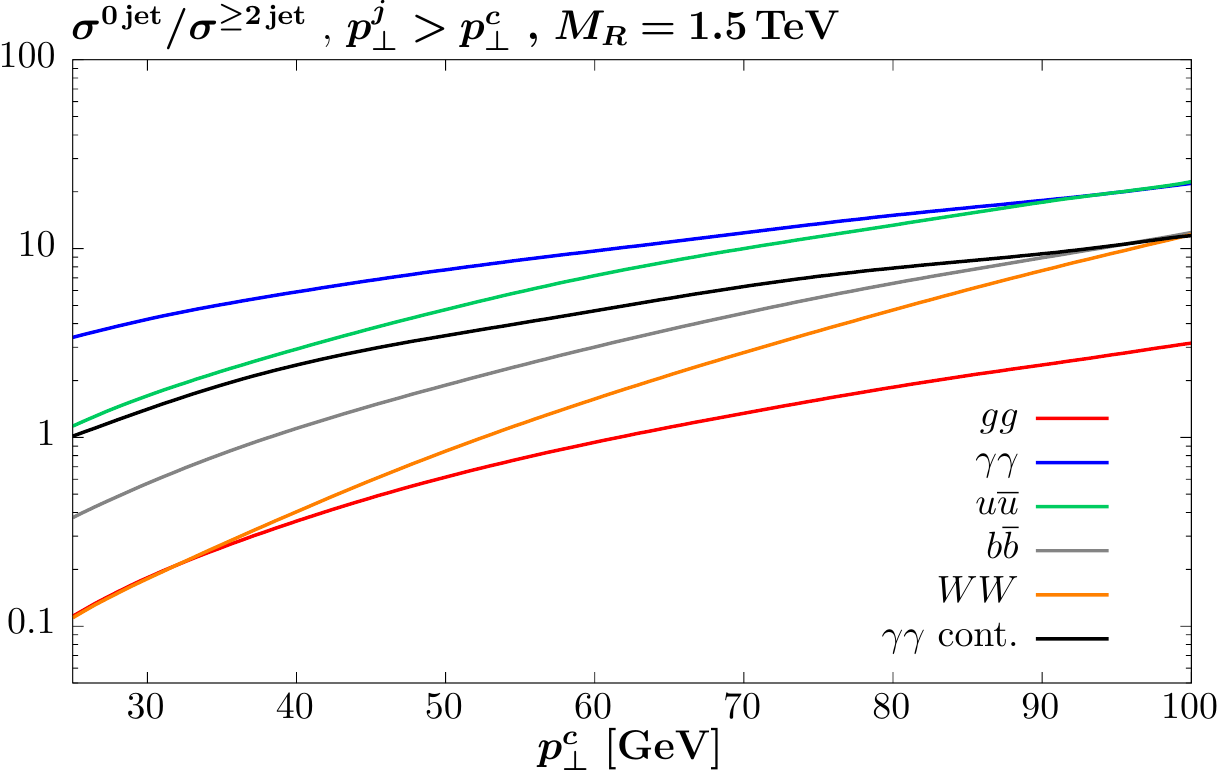}
\includegraphics[scale=0.65]{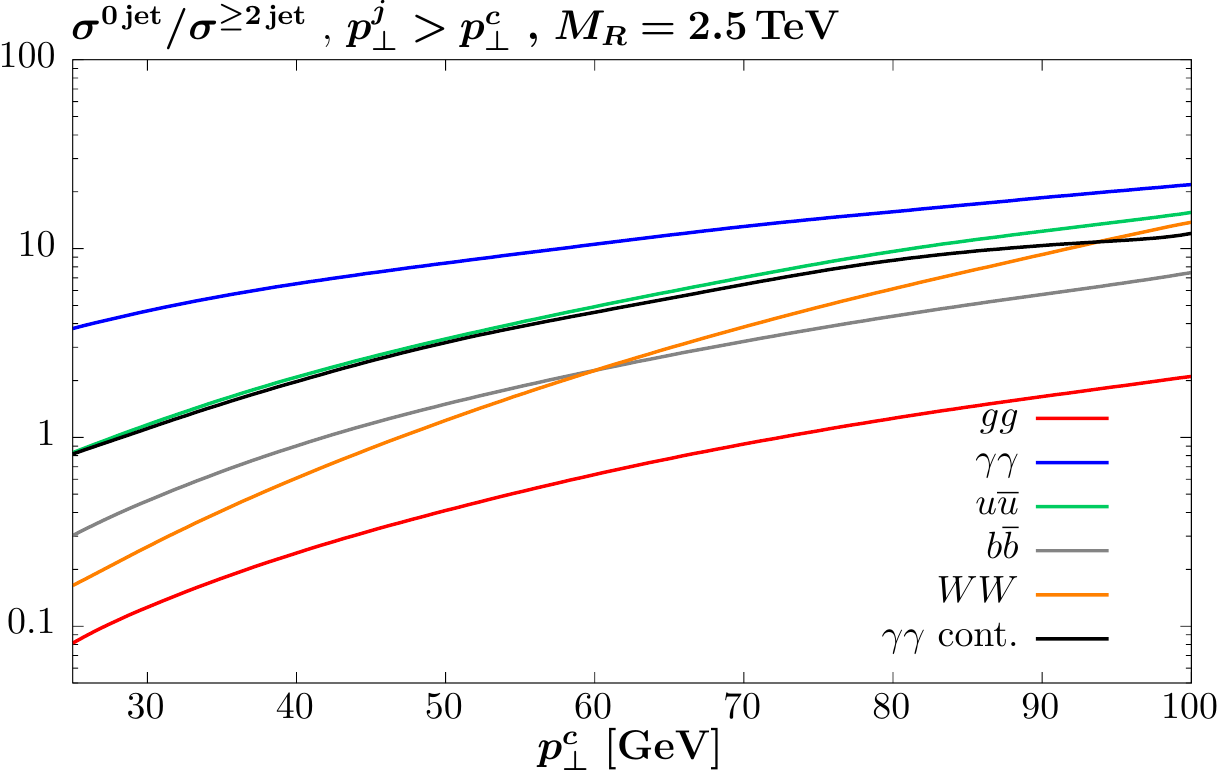}
\caption{Ratio of 0 jet to $\geq$ 2 jet cross sections as a function of the minimum jet $p_\perp$ for a range of initial--state resonance production processes, and for the continuum $\gamma\gamma$ background. Results shown for resonance masses $M_R=0.75,\,1.5,\,2.5$ TeV.}
\label{fig:jetptveto1}
\end{center}
\end{figure} 

We also show in Figs.~\ref{fig:jetmultiq} and~\ref{fig:jetmultiqpt50}  the predictions for heavy $b\overline{b}$ and light $u\overline{u}$ quark mediated production. The results for the three mass points are similar, with some increase in jet multiplicity with increasing $M_R$, similar to the $gg$ case. We observe a non--negligible difference in the distributions between the heavy and light quark cases, with the $b\overline{b}$ distribution following the $gg$ closely, while the $u\overline{u}$ is expected to be accompanied by a lower jet multiplicity. The difference between the $u\overline{u}$ and $gg$ cases is in line with QCD expectations, for which the average particle multiplicity from an initial--state gluon is higher than for a quark~\cite{Dokshitzer:1991wu}. For the $b\overline{b}$ case, as the $b$--quark PDF is generated entirely by DGLAP emission above the $b$--quark threshold, there is a much greater contribution from the $g\to b\overline{b}$ process in the initial--state, where the outgoing $b$--jet is observed within the jet acceptance. Therefore, by requiring one or more $b$--jets in the final state, the $b\overline{b}$ initiated production mode may be identified or constrained. In Fig.~\ref{fig:bjetmulti} we show the $b$--jet multiplicity in this production mode for a range of jet $p_\perp$ cuts, and we can see that roughly $\sim 50\%$ of events are expected to be accompanied by an additional $b$--jet in the final state (for other initial states this fraction is of course significantly lower). The case of $M_R=0.75$ TeV is shown for concreteness, although similar results hold for the other mass points.

Although Figs.~\ref{fig:jetmulti}--\ref{fig:jetmultiqpt50} only consider two choices of jet cut, we can also generalise this, showing in Fig.~\ref{fig:jetptveto} the ratio of 0 to $\geq 1$ jet cross sections as a function of the minimum jet $p_\perp$. The same hierarchy in jet ratios between the different production modes and the same trends with increasing $M_R$ described above are clear. Such distributions have been considered in~\cite{Ebert:2016idf} within a distinct analytic SCET approach, for all cases but the $\gamma\gamma$ and $WW$ initial states, and the results are found to be similar. In addition, in~\cite{Harland-Lang:2016qjy} a simple analytic approach is taken to calculate the 0--jet cross sections in $gg$ and $\gamma\gamma$--mediated production, and again the results are very similar to those here for these cases.

It is interesting to observe that the expected trend with  the jet $p_\perp$ cut in Fig.~\ref{fig:jetptveto} is also in general different between the various production modes. In particular, as the $p_\perp$ cut is increased the jet ratio in the quark mediated processes increases more rapidly, such that for $p_\perp \gtrsim 60 $ GeV the $u\overline{u}$ ratio is in fact higher than the $\gamma\gamma$. On the other hand, while for lower jet $p_\perp$ cut, the $\gamma\gamma$ continuum ratios are very similar to the $u\overline{u}$, as expected from the dominantly light $q\overline{q}$ initial--state for this process, as the $p_\perp$ cut is increased, and the jet ratios become more sensitive to the structure of the production process, this is no longer the case, with the jet multiplicity being higher in the continuum case. This indicates that at these higher $p_\perp$ values the predictions may be sensitive to the precise nature of the resonance production process, which is not included in the effective description considered here. 

Finally, in Fig.~\ref{fig:jetptveto1} we show the 0 to $\geq 2$ jet ratios. The relative hierarchy and trends with the jet $p_\perp$ cut are comparable to the 0 to $\geq 1$ case, but in fact the separation between the $\gamma\gamma$ and $gg$ predictions, for example, is increased. This indicates that a greater discrimination can be achieved by considering the full jet multiplicity distribution as in Fig.~\ref{fig:jetmulti} rather than an individual ratio. In the following section we will consider quantitively what limits may be set on the different production modes using such distributions.

\section{Comparison to ATLAS data}\label{sec:ATLAS}

\begin{figure}[h]
\begin{center}
\includegraphics[scale=0.65]{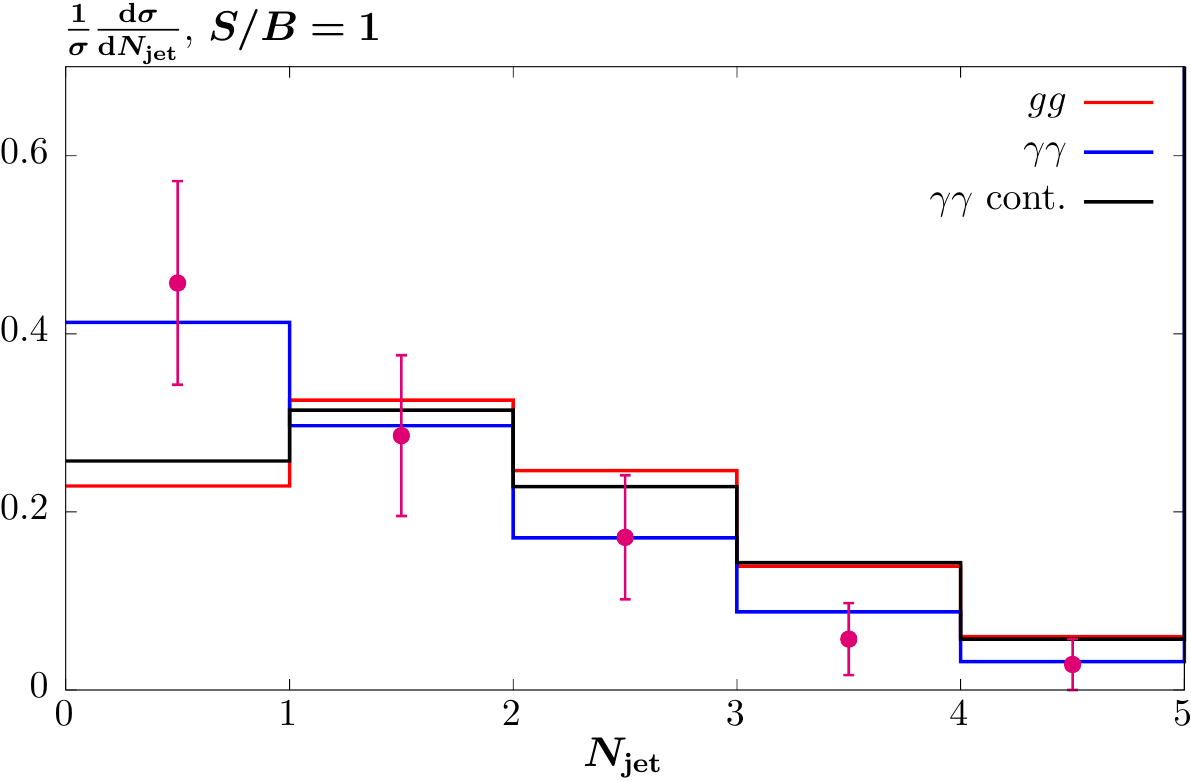}
\includegraphics[scale=0.65]{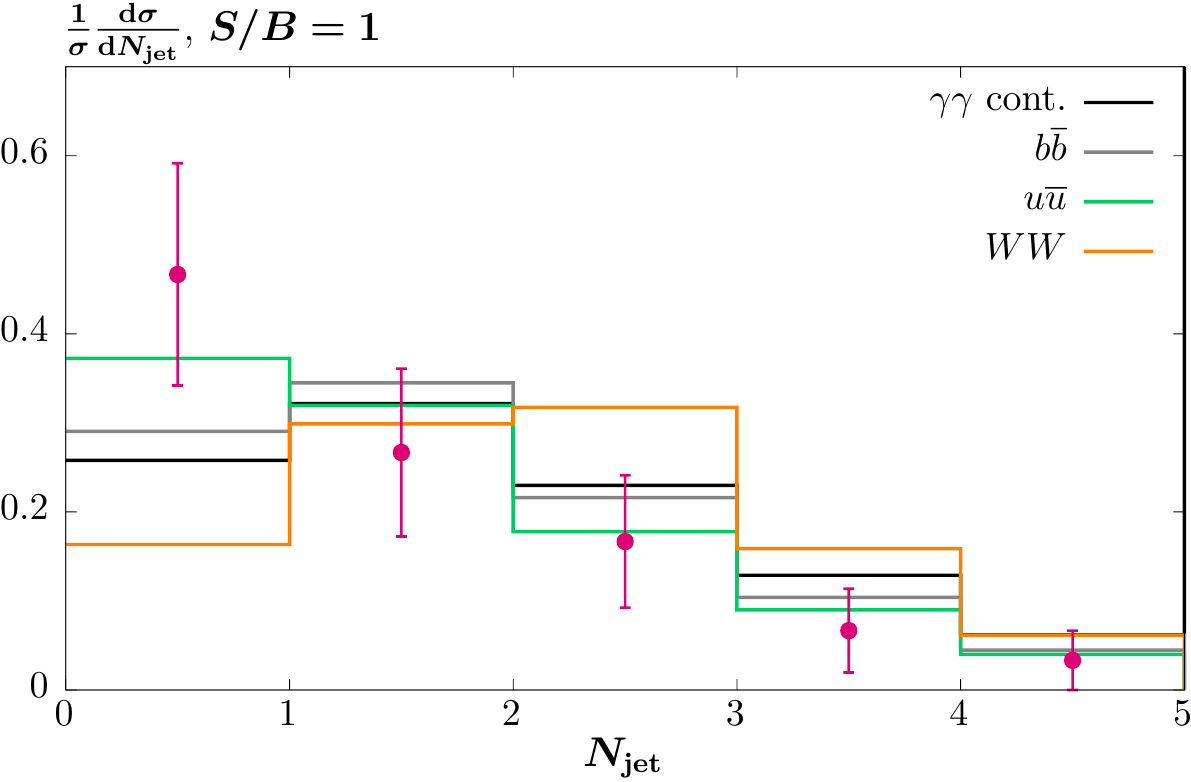}
\caption{Exclusive jet multiplicities, for different initial--state resonance production processes, and the SM continuum $\gamma\gamma$ production process, compared to ATLAS~\cite{Aaboud:2016tru}  measurement in the range $700<M_{\gamma\gamma}<840$ GeV. The continuum background is taken from~\cite{Aaboud:2016tru}, and is included in all signal distributions, assuming a $S/B$ ratio of 1.}
\label{fig:jetmultiat}
\end{center}
\end{figure} 

In the ATLAS analysis~\cite{Aaboud:2016tru} a measurement is presented of the exclusive jet multiplicity in the signal $700<M_{\gamma\gamma}<840$ GeV region for the spin--0 resonance selection described in Section~\ref{sec:analysis}. Although no significant excess is observed in the larger 2016 data set, it is nonetheless instructive to compare the results of our study with these data. It should however be emphasised that such data have limited statistics  and moreover as these are not presented in an unfolded form, our comparison can only be approximate, as it omits a full ATLAS detector simulation as well as the inclusion of pile--up (although pile--up jets are largely rejected by cuts based on tracking information). Nonetheless for these data the dominant source of experimental uncertainty is certainly statistical, and moreover the more significant differences predicted between for example the $\gamma\gamma$ and $gg$ are visible even in light of this, as we will show. In our comparison we include a contribution from the SM continuum $\gamma\gamma$ background, which to be as close to the ATLAS analysis as possible, we simply take from their quoted \texttt{SHERPA}~\cite{Gleisberg:2008ta} MC predictions for the $\gamma\gamma$ continuum (which represent 90\% of the total background). We take a $S/B$ ratio of 1, roughly corresponding to that seen in the data, and compare to the $N_{\rm jet}\leq 3$ bins, where 29 of the 31 observed events lie, and the theoretical predictions are more reliable.

\begin{table}
\begin{center}
\def\arraystretch{1.4}
\begin{tabular}{|c|c|c|c|c|c|c|}
\hline
Channel & $\gamma\gamma$& $gg$& $b\overline{b}$&$u\overline{u}$&$WW$&$\gamma\gamma$ cont. \\ \hline
$\chi^2$&0.4&7.1&3.9&1.4&14.1&5.6\\
\hline
\end{tabular}
\caption{The $\chi^2$ values for the description of the four $N_{\rm jet} \leq 3$ bins for the ATLAS~\cite{Aaboud:2016tru} measurement of the exclusive jet multiplicities, for different initial--state resonance production processes. The contribution from the SM $\gamma\gamma$ continuum, taken from~\cite{Aaboud:2016tru}, is included, and $S/B=1$ is assumed. These values correspond to the distributions shown in Fig.~\ref{fig:jetmultiat}.}
\label{table:jetmultiat}
\end{center}
\end{table}

The predicted distributions for the production processes discussed above are shown in Fig.~\ref{fig:jetmultiat}, along with the ATLAS data. From a straightforward visual comparison, it is clear that the data show some preference for the $\gamma\gamma$ and $u\overline{u}$ scenarios, in comparison to the background only and in particular the $gg$ and $WW$ cases. To give a more precise estimate, we can evaluate the corresponding $\chi^2$ values\footnote{Excluding the $N_{\rm jet}=3$ bin from the comparison, for which the observed 2 events is quite low, does not significantly affect the results which follow.}, as shown in Table~\ref{table:jetmultiat}. These results confirm the above expectation, with the description in the $gg$ case being poor; if we assume $n_{\rm dof}=n_{\rm bins}-1=3$ for these normalised distributions, this corresponds to roughly a $\sim 95\%$ exclusion. On the other hand, the description is very good in the $\gamma\gamma$ and $u\overline{u}$ cases\footnote{The fact that the description of the data is so good in the $\gamma\gamma$ case, giving a $\chi^2/{\rm dof}\ll 1$, should clearly not be expected, and indeed for other choices of $S/B$ and merging scale $Q_{\rm cut}$, for example, the value can be higher, although nonetheless corresponding to a very good description of the data. Another contributing factor is that the  errors  on the ATLAS data are somewhat overestimated for this normalised observable: see below.}. Interestingly, the description for the pure $\gamma\gamma$ background hypothesis is also relatively poor. Finally, the description for a purely $WW$ initial state is very poor indeed, corresponding to a $\sim 99.8\%$ exclusion; we expect similar results in the $ZZ$ and, to a lesser extent, $\gamma Z$ modes.

The description in the $b\overline{b}$ case is reasonable, but we note that here by far the most discriminating variable is, as discussed in the previous section, the $b$--jet multiplicity. Even with the relatively low statistics data sample, a measurement of this observable would lend strong support, or place strong constraints on, such a scenario. Indeed, in~\cite{Aaboud:2016tru} it is reported that, with a $b$--tagging efficiency of about 85\%, roughly 8\% of events in the signal region are found to contain $b$--jets, consistent within statistical uncertainties with the sideband regions. Comparing with Fig.~\ref{fig:bjetmulti}, we can see that $\sim 60\%$ of signal events in the purely $bb$--initiated scenario are predicted to contain $b$--jets in the ATLAS event selection. Even accounting for the continuum background such a result is in strong tension with a dominantly $bb$--induced production mode, although to evaluate the exact constraints would require a more precise comparison accounting for  the $b$--jet efficiency and mis--ID rate. We will therefore not consider the $b\overline{b}$ case in what follows.

While the precise values of the $\chi^2$ depend on the uncertain $S/B$ ratio, and merging scale $Q_{\rm cut}$, these are only found to vary by $\sim 10\%$ for reasonable changes in these parameters. Taking our simulation for the $\gamma\gamma$ continuum background leads to somewhat larger values of $\chi^2$, although still with the $\gamma\gamma$ and $u\overline{u}$ cases giving very good descriptions of the data, with the same hierarchy observed for the other scenarios. 

Before concluding this section, it is worth noting that the transverse momentum $p_\perp^{\gamma\gamma}$ of the diphoton system has also been measured in~\cite{Aaboud:2016tru}, where it is found that $\sim 60\%$ of the observable cross section has $p_\perp^{\gamma\gamma}<60$ GeV. Such an observable is also sensitive to the production mode of the resonance, with the $gg$ and in particular $WW$ hypotheses predicting broader distributions. While the description of the measured $p_\perp^{\gamma\gamma}$ distribution is in fact relatively poor in all cases, including the background only hypothesis, due to the apparent downward fluctuation in the second bin, the same hierarchy in the quality of the data descriptions  is found  as in the case of the jet multiplicity. While these two observables are evidently correlated, this nonetheless gives a consistency check of the overall approach, and indeed with further data and finer binning, the $p_\perp^{\gamma\gamma}$ observable may also allow further discrimination between the modes.

Finally, it is worth emphasising that the results presented above are from a statistical point of view expected to be conservative. In particular, the statistical errors on the normalised jet multiplicity distribution presented by ATLAS in~\cite{Aaboud:2016tru} are simply found by rescaling the errors on the observed event numbers in each bin by the total number of observed events. However this ignores the positive correlation between the measured numbers of events in each bin with the total number of observed events, which will lead in general to a reduction in the error size as well as a correlation between different bins for the normalised distribution; for the relatively low number of measured events this effect will not necessarily be negligible. Indeed, taking a simplified approach and treating all uncertainties on the normalised distribution as Gaussian, we find a that a more complete treatment including the full correlation matrix between the different bins leads to a further deterioration in the quality of the description for the $gg$, $WW$ and background only scenarios. However, to be conservative we do not present such a comparison in detail here.

\section{Expected limits on production modes}\label{sec:exc}

 \begin{figure}
\begin{center}
\includegraphics[scale=0.6]{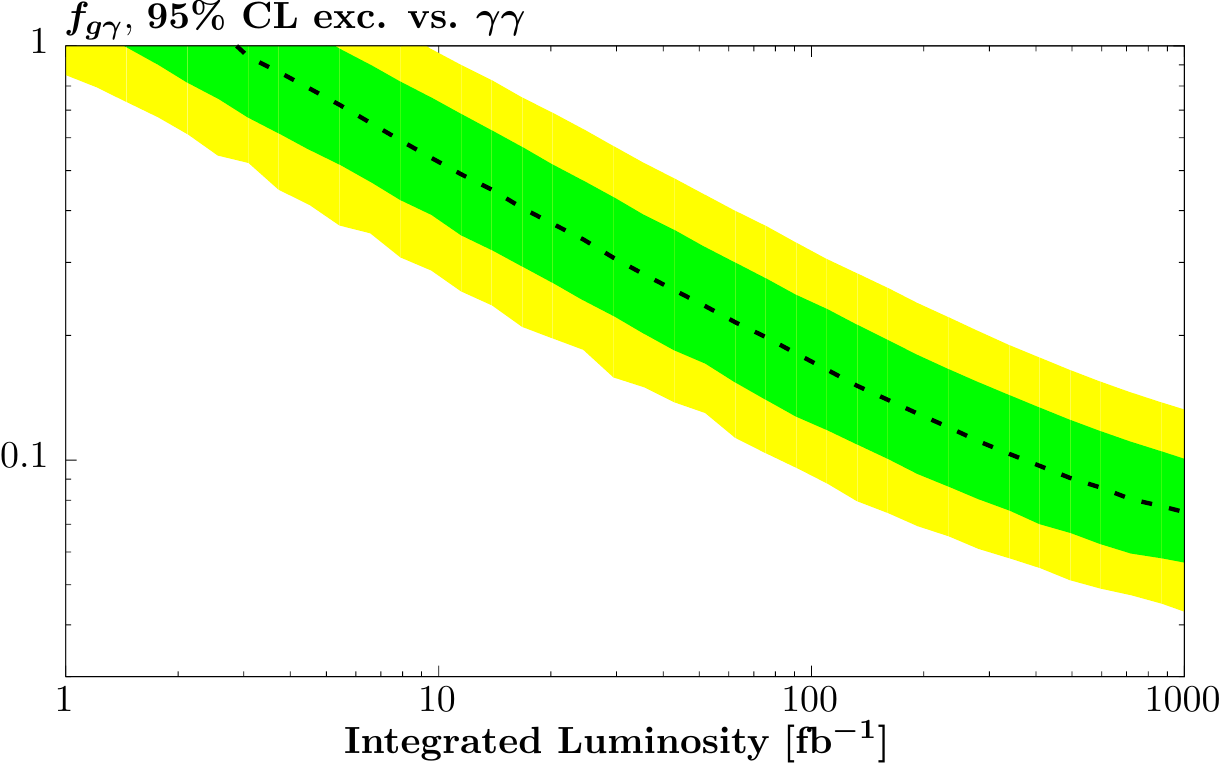}
\includegraphics[scale=0.6]{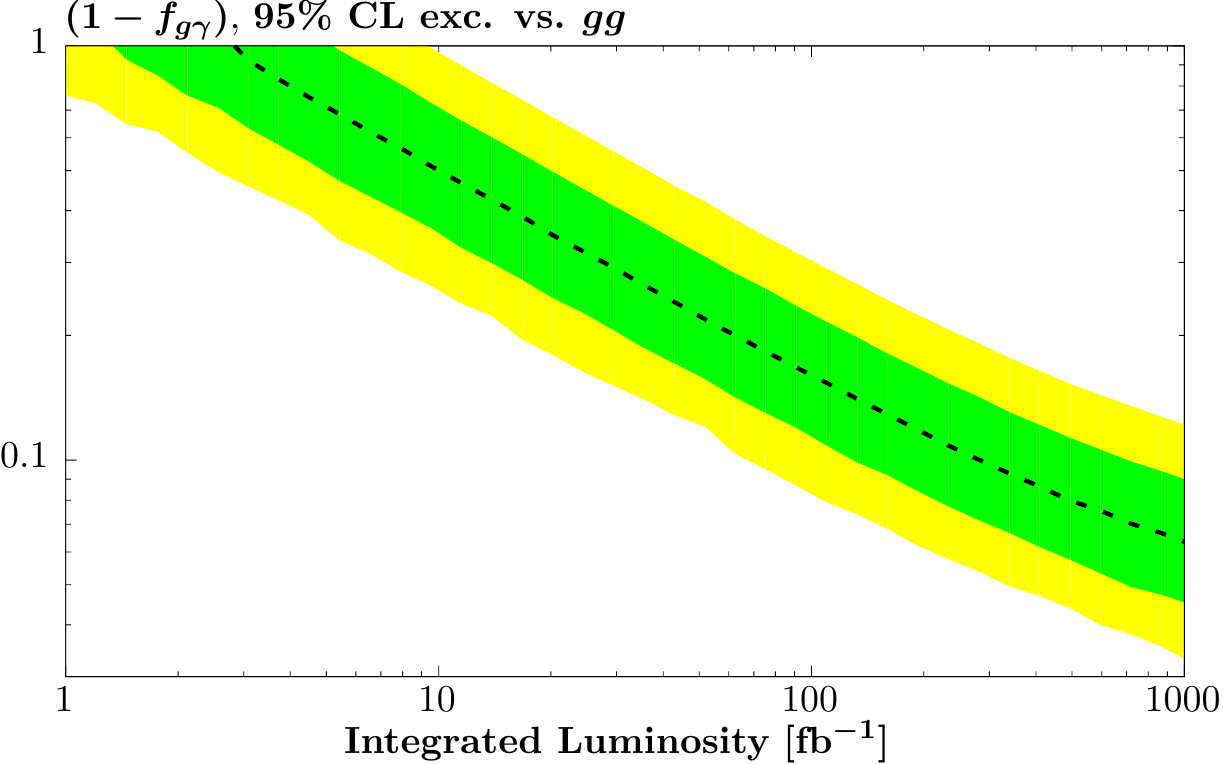}\par
\vspace{0.3cm}
\includegraphics[scale=0.6]{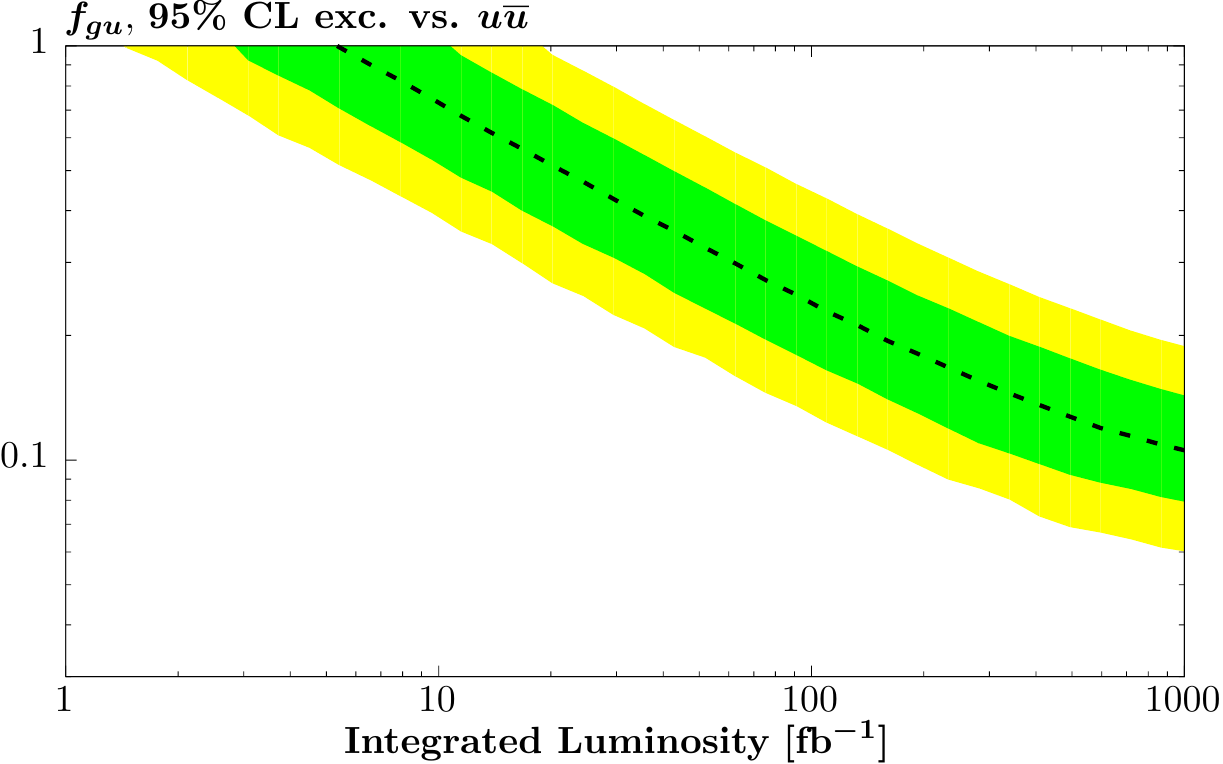}
\includegraphics[scale=0.6]{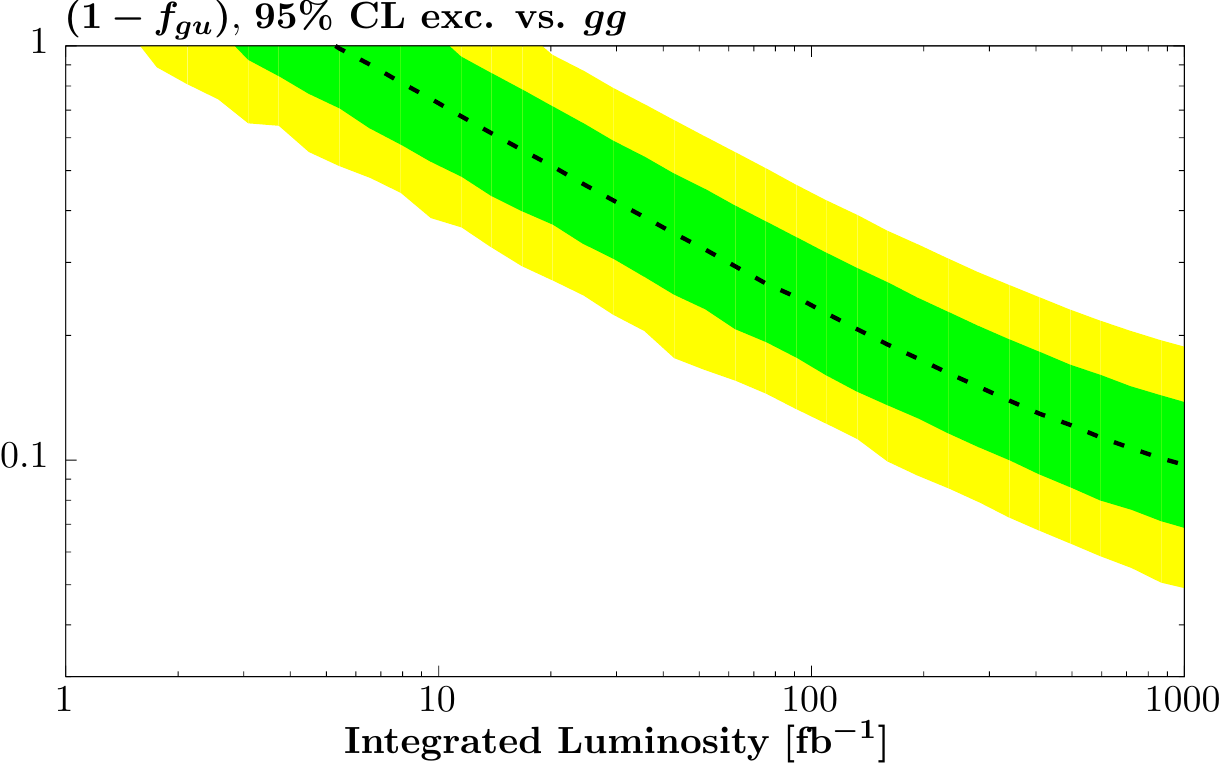}\par
\vspace{0.3cm}
\includegraphics[scale=0.6]{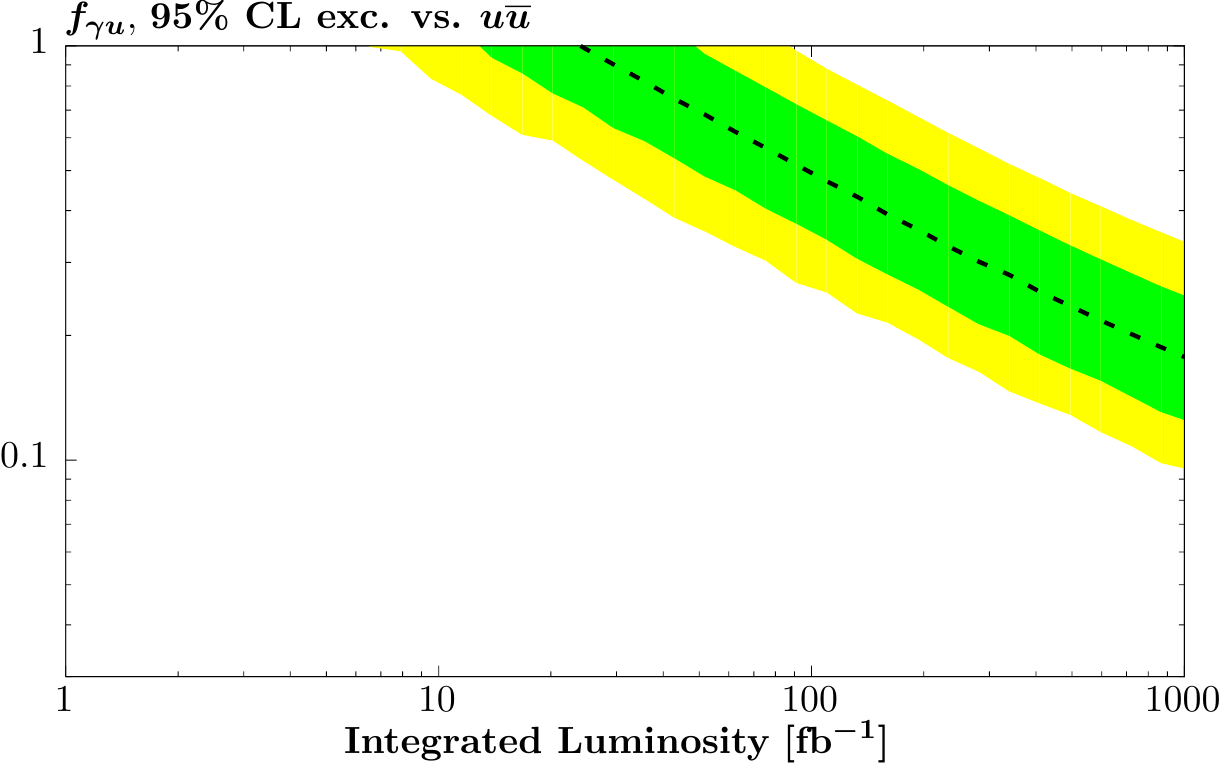}
\includegraphics[scale=0.6]{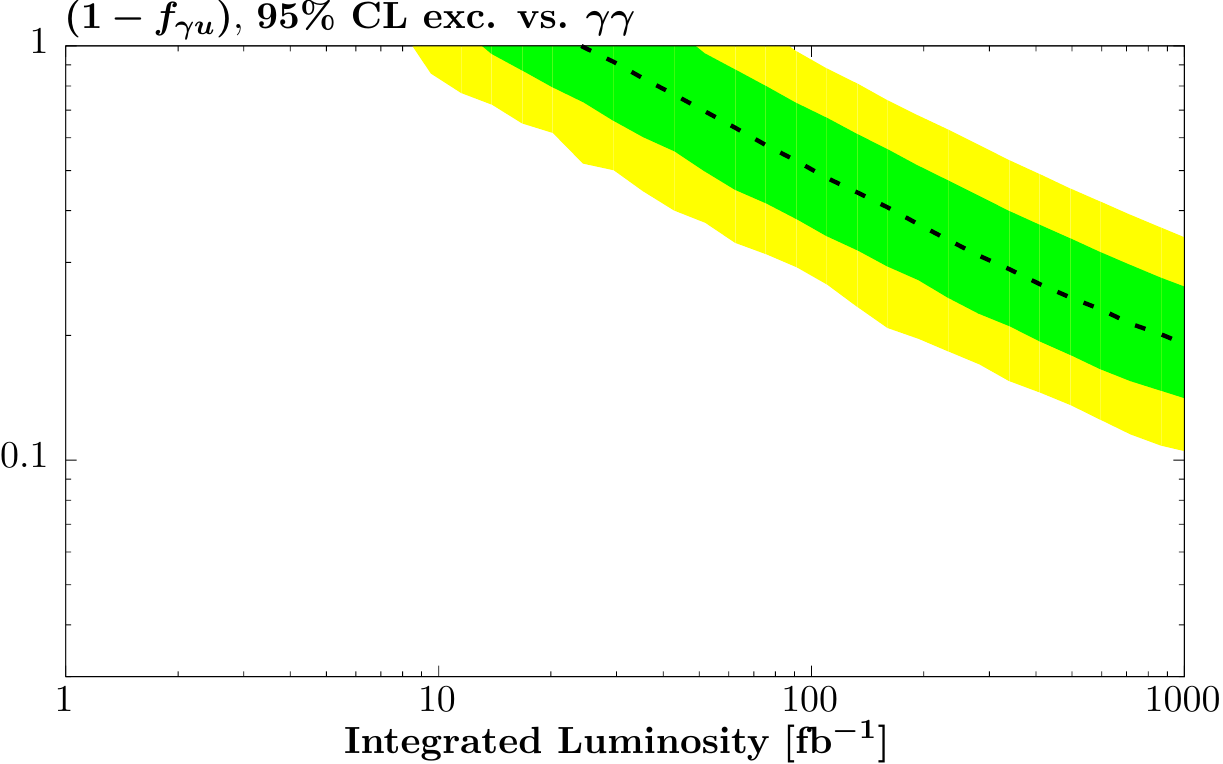}
\caption{The expected 95 \% exclusion limits on the fractional contributions to a $M_R=0.75$ TeV scalar resonance production cross section, described in the text, against the hypotheses of pure $gg$, $\gamma\gamma$ or $u\overline{u}$--initiated production. For example, in the upper two plots, $f_{g\gamma}=1(0)$ corresponds to purely $gg$($\gamma\gamma$) induced production, and similarly for the other plots. The dashed lines correspond to the median value of the limit, and the $\pm 1$ and 2 $\sigma$ ranges are shown.}
\label{fig:CLlim}
\end{center}
\end{figure} 

Having found such promising results when comparing to the  limited 2015 ATLAS jet multiplicity data, it is natural to consider what limits we can expect on the production modes. To be more precise, we will perform a confidence limit ($CL$) hypothesis test, analogous to those used by the ATLAS~\cite{Aad:2015mxa} and CMS~\cite{Khachatryan:2014kca} analyses for determining the spin and parity of the SM Higgs. The test statistic is defined as
\begin{equation}
Q=-2\ln \frac{\mathcal{L}(h_1)}{\mathcal{L}(h_2)}\;,
\end{equation}
for the initial--state hypothesis, i.e. $h=\gamma\gamma, gg, q\overline{q}$ or more generally some non--trivial admixture of these. This can be used to discriminate between an initial--state production hypotheses $h_1$ in favour of $h_2$. For an observed value of $Q_{\rm obs}$ the exclusion of the $h_2$ in favour of $h_1$ is given in terms of the modified confidence level
\begin{equation}
{\rm CL}_s=\frac{P(Q\geq Q_{\rm obs}|h_1)}{P(Q\geq Q_{\rm obs}|h_2)}
\end{equation}
where $P(Q\geq Q_{\rm obs}|h)$ is the probability that the test statistic is at least as high as $Q_{\rm obs}$ under a hypothesis $h$. 

We can thus use this approach to calculate the expected exclusion limits on a specified production mode $h_2$, assuming the resonance is produced  via $h_1$. To be more concrete we assume three simplified scenarios where the produced resonance is only produced via two of the three $\gamma\gamma$, $gg$ and $u\overline{u}$ initial--states, i.e.
\begin{align}\nonumber
\sigma_{gg}&=f_{g\gamma}\, \sigma_{\rm R} \quad\quad \sigma_{\gamma\gamma}=(1-f_{g\gamma})\,\sigma_{\rm R}\;,\\ \nonumber
\sigma_{gg}&=f_{g u}\, \sigma_{\rm R} \quad\quad \sigma_{u\overline{u}}=(1-f_{g u})\,\sigma_{\rm R}\;, \\ \label{fdef}
\sigma_{\gamma\gamma}&=f_{\gamma u}\, \sigma_{\rm R} \quad\quad \sigma_{u\overline{u}}=(1-f_{\gamma u})\,\sigma_{\rm R}\;,
\end{align}
where $\sigma_R$ is the signal cross section; we do not consider the $WW$ initial state for simplicity, but will comment on this below. We take the case of $M_R=0.75$ TeV in what follows, but similar results follow for higher mass points. To see how these limits can be expected to extend given a certain amount of luminosity, we will for concreteness consider an observed cross section of 7 fb after cuts in the narrower $725<M_{\gamma\gamma}<775$ GeV region, with the SM continuum (taken now from our MC sample) and resonance signal present equally, i.e. $S/B=1$. We also include a 10\% systematic uncertainty on the data, as a fairly conservative assumption.

The expected 95\% exclusion limits on the cross section fractions in (\ref{fdef}) are shown in Fig.~\ref{fig:CLlim}. In each case we consider the limit on the fraction $f$ against the hypotheses that the resonance is purely produced in the corresponding production modes; for example we calculate the exclusion on $f_{g\gamma}$ under both the purely $gg$ and $\gamma\gamma$ hypotheses. We can see that already for the integrated luminosity of $3.2$ ${\rm fb}^{-1}$ corresponding to the ATLAS data, a 95\% exclusion of the purely $gg$ hypotheses, corresponding to $f_{g\gamma},f_{gu}=1$, is consistent with the projected limits, and with the results of the previous section. Indeed, if we treat the fractions $f$ in (\ref{fdef}) as free parameters and perform a $\chi^2$ minimisation and profile test with respect to the ATLAS data we find that $f_{g\gamma} (f_{g u}) < 0.71 (0.76)$ at 95\% confidence, with minima at 0, while the $f_{\gamma u}$ is unconstrained; these results are again completely consistent with these expected confidence limits. We can moreover see that with just a few $10$s of ${\rm fb}^{-1}$ any significant $gg$--induced component can be excluded in favour of the purely $\gamma\gamma$ and $u\overline{u}$ hypotheses, and conversely any significant $\gamma\gamma$ or $u\overline{u}$ induced component can be excluded in favour of the purely $gg$. 

On the other hand, for the $u\overline{u}$ and $\gamma\gamma$ scenario, for which the two available production modes predict closer jet multiplicity distributions, see Fig.~\ref{fig:jetmultiat}, the situation is more challenging, and a larger $O(100)$ ${\rm fb}^{-1}$ sample is required for a sizeable contribution from the $\gamma\gamma$ mode to be excluded in favour of the purely $u\overline{u}$, and vice versa. In such a situation it is likely that other methods for distinguishing between the different initial--states will be more competitive. For example, as discussed in~\cite{Fichet:2015vvy,Harland-Lang:2016qjy}, a measurement of just a few resonance events in the essentially background free exclusive channel, where both protons remain intact after the collision, will provide strong evidence in favour of a significant contribution from the $\gamma\gamma$ mode, and conversely any lack of observed events will disfavour this. Alternatively, by selecting events with rapidity gaps in the final state the $\gamma\gamma$--initiated contribution may be isolated: it is worth recalling, in particular, that a significant fraction of $\gamma\gamma$--initiated events are expected to occur due to low--scale coherent emission from the protons, which naturally lead to rapidity gaps in the final--state, see~\cite{Harland-Lang:2016apc}.

Finally, returning to the VBF channel, if we consider the simplified scenario that the resonance is produced by the $WW$ and $\gamma\gamma$ modes, i.e.
\begin{equation}
\sigma_{WW}=f_{W\gamma}\, \sigma_{\rm R} \quad\quad \sigma_{\gamma\gamma}=(1-f_{W\gamma})\,\sigma_{\rm R}\;,
\end{equation}
then we already find that $f_{W\gamma}<0.48$ at 95\% confidence, with similar limits if the additional mode is $gg$ or $u\overline{u}$. 

\section{Conclusion}\label{sec:conc}

The observation by the ATLAS and CMS collaborations, in roughly $3\,{\rm fb}^{-1}$ of data at $\sqrt{s}=13$ TeV recorded in 2015, of an excess of events around 750 GeV in the diphoton mass spectrum provoked a great deal of theoretical interest. Although no significant excess was seen in the increased 2016 data set, this initial observation has motivated us to  discuss in detail the possibility of using measurements of the jet multiplicity associated with the production of a high--mass resonance as a means to distinguish between different production mechanisms.

Specifically, we have considered the $\gamma\gamma$, $gg$, $WW$, light quark $q\overline{q}$ and heavy $b\overline{b}$ quark initiated processes and shown that in each case the predicted level of jet activity is quite different. In particular, due to the small size of $\alpha$ and the lower photon branching probability, the jet multiplicity is expected to be lower in the $\gamma\gamma$ case compared to the QCD quark and gluons. As the particle multiplicity associated with an initial--state gluon is higher compared to the quark case, the higher jet activity in the $gg$ case also allows this mechanism to be separated from the $q\overline{q}$. For the $WW$--initiated channel the resonance is generally produced in association with at least two additional jets, due to the relatively high $p_\perp$ recoiling quarks in the final state. For the heavy $b\overline{b}$ the most discriminating observable is instead simply the $b$--jet fraction, which is predicted to be significantly higher than in all other scenarios.

To demonstrate how such an approach may be applied to data, in this paper we have compared these results with the 2015 ATLAS measurement of the jet multiplicity in the spin--0 signal region associated with the initial diphoton excess observation. We have found that even with these fairly limited data, a purely $gg$--initiated scenario is disfavoured, while the light $q\overline{q}$ and $\gamma\gamma$ scenarios provide a good description. The continuum background--only hypothesis also gives a somewhat worse description than these latter cases. In addition, we have found that a dominantly $WW$--initiated production mechanism is in strong tension with the data, and we have seen that the relatively small observed $b$--jet fraction disfavours a dominantly $b\overline{b}$--initiated production mechanism. We have also presented expected exclusion limits on different production hypotheses with the collected integrated luminosity. With just a few $10$'s of ${\rm fb}^{-1}$, we can expect to rule out any sizeable fraction of $gg$ in favour of light $q\overline{q}$ and $\gamma\gamma$ production, and vice versa. It is more challenging to distinguish between the $u\overline{u}$ and $\gamma\gamma$ modes, although still possible with enough data.

If a high mass resonance is observed at the LHC or a future collider, one of the first tasks will be to determine as precisely as possible the nature of such a new state. It has been the goal of this paper to present a method for distinguishing between various production modes, which can be applied to any heavy resonance which we might hope to find in the future.

\section*{Acknowledgements}

We thank Michelangelo Mangano, Josh McFayden, Valya Khoze, Frank Krauss and Robert Thorne for useful discussions. VAK thanks the Leverhulme Trust for an Emeritus Fellowship. The work of MGR  was supported by the RSCF grant 14-22-00281. LHL thanks the Science and Technology Facilities Council (STFC) for support via the grant award ST/L000377/1. MGR thanks the IPPP at the University of Durham for hospitality. MS is supported in part by the European Commission through the ``HiggsTools'' Initial Training Network PITN-GA-2012- 316704. 

\appendix

\section{Cross check: comparison to spin--2 selection}\label{sec:ATLAS2}

In addition to the ATLAS spin--0 selection data considered in Section~\ref{sec:ATLAS}, a spin--2 event selection was also taken in~\cite{Aaboud:2016tru}, identical to the spin--0 case, but with a lower $E_\perp>55$ GeV cut on the final--state photons. Thus the corresponding number of events in this case is larger, with the sample in the spin--0 case being a subset of the spin--2; in the signal $700 < M_{\gamma\gamma}<840$ GeV region ATLAS find 31 (70) events for the spin--0 (2) samples. However, if the excess of events is indeed due to a scalar resonance decaying (isotropically) to photons then the lower $E_\perp$ cuts of the spin--2 selection, for which the relative contribution from the continuum background will be larger, are expected to lead to an overall decrease in the $S/B$ ratio.

\begin{figure}[h]
\begin{center}
\includegraphics[scale=0.65]{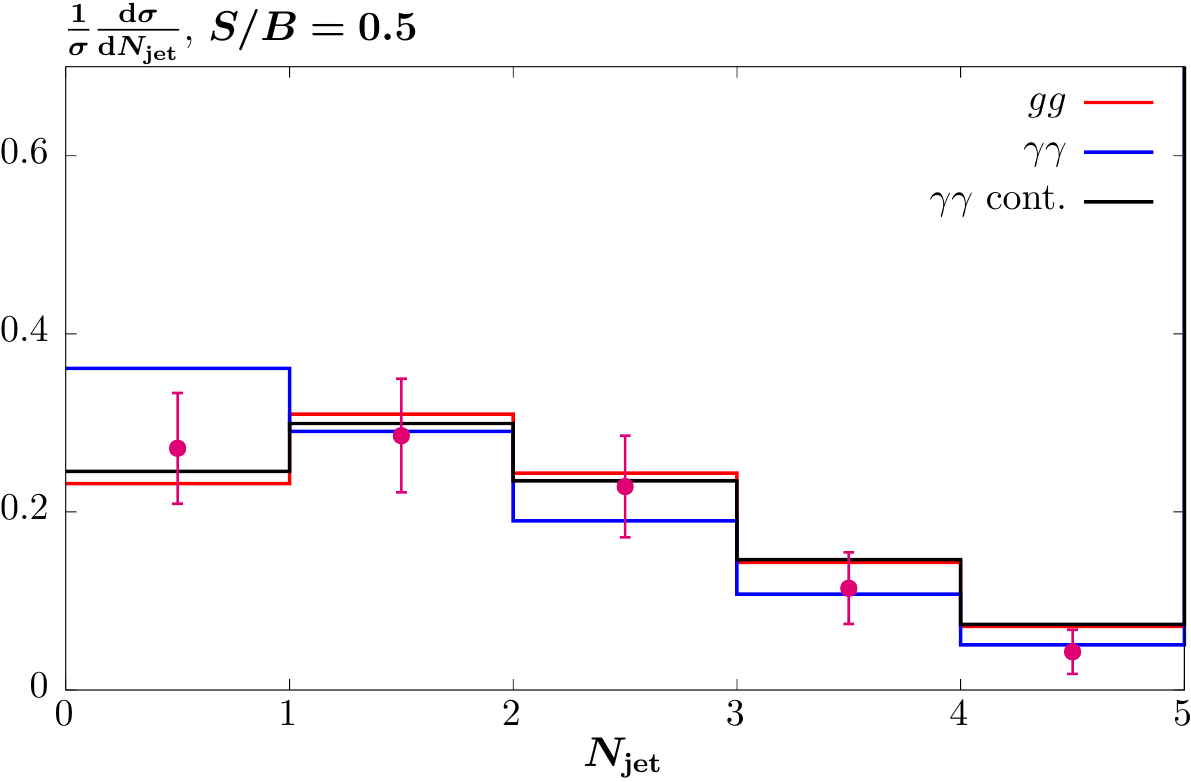}
\includegraphics[scale=0.65]{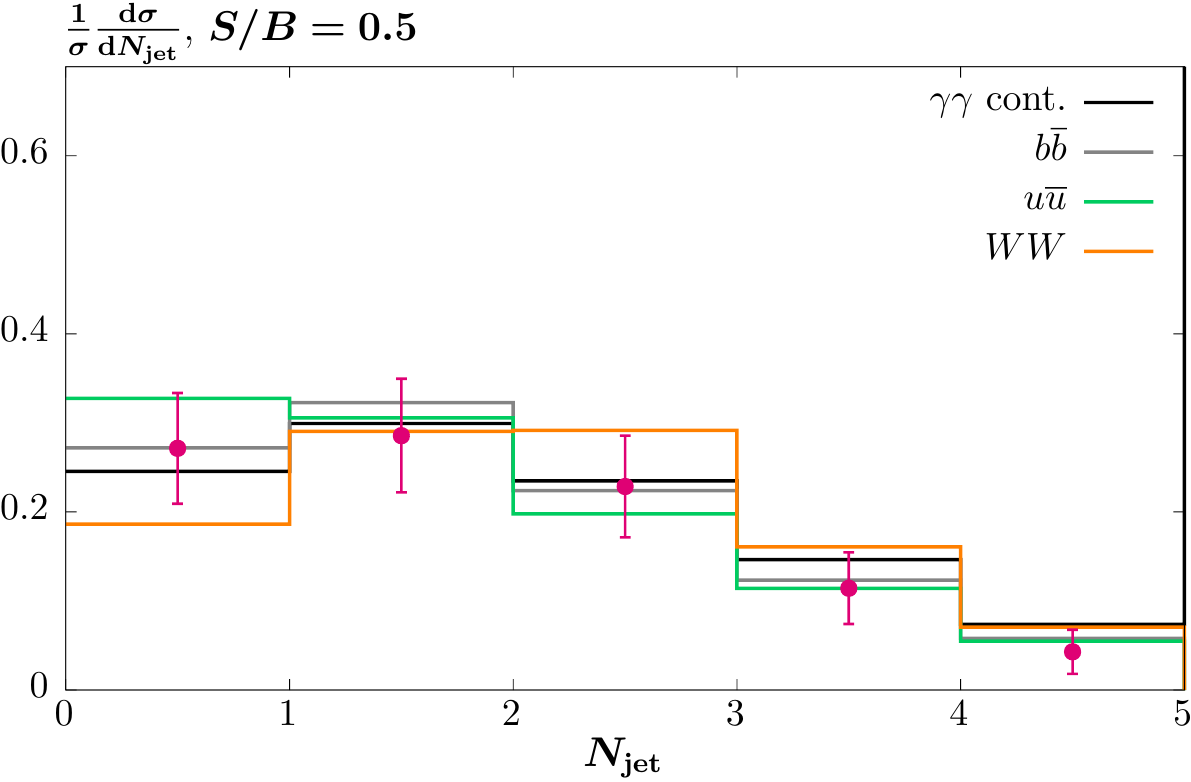}
\caption{Exclusive jet multiplicities, as in Fig.~\ref{fig:jetmultiat}, but compared to data taken with the ATLAS~\cite{Aaboud:2016tru}  spin--2 event selection. The continuum background is taken from~\cite{Aaboud:2016tru}, and is included in all signal distributions, assuming a $S/B$ ratio of 0.5.}
\label{fig:jetmultiats2}
\end{center}
\end{figure}

Nonetheless, it is a useful cross check to compare the expected jet multiplicity distributions with the higher statistics data corresponding to the spin--2 selection, still taking our spin--0 resonance hypothesis. We estimate from~\cite{Aaboud:2016tru} that $S/B\sim 0.5$, lower than the $S/B\sim 1$ found for the spin--0 selection, and consistent with the discussion above, and moreover with a $S/\sqrt{B}$ sensitivity that is comparable in both cases, as is found in the ATLAS analysis. The results are shown in Fig.~\ref{fig:jetmultiats2}: due to the larger background contribution, we can see that the different signal distributions are less well separated, although the smaller statistical errors on the data might in principle allow an increased differentiation. However, we can see from a rough visual comparison that all hypotheses appear to provide a relatively good description of the data, although with the prediction in the 0--jet bin lying a little above the data in the $\gamma\gamma$ case. Nonetheless, in Table~\ref{table:jetmultiats2} we show the corresponding $\chi^2$ values and we can see that description of the data is good in all cases, with a $\chi^2/{\rm dof}\lesssim 1$ for all hypotheses; this is driven by the fact that the continuum background, which gives the dominant contribution to the measured sample, provides an excellent description of the data.

\begin{table}
\begin{center}
\def\arraystretch{1.4}
\begin{tabular}{|c|c|c|c|c|c|c|}
\hline
Channel & $\gamma\gamma$& $gg$& $b\overline{b}$&$u\overline{u}$&$WW$&$\gamma\gamma$ cont. \\ \hline
$\chi^2$&2.2&1.0&0.2&1.0&4.0&0.8\\
\hline
\end{tabular}
\caption{The $\chi^2$ values for the description of the four $N_{\rm jet} \leq 3$ bins for the ATLAS~\cite{Aaboud:2016tru} measurement of the exclusive jet multiplicities, for different initial--state resonance production processes. The contribution from the SM $\gamma\gamma$ continuum, taken from~\cite{Aaboud:2016tru}, is included, and $S/B=0.5$ is assumed. These values correspond to the distributions shown in Fig.~\ref{fig:jetmultiats2}.}
\label{table:jetmultiats2}
\end{center}
\end{table}

\bibliography{references}{}
\bibliographystyle{h-physrev}

\end{document}